\documentclass[fleqn,usenatbib]{mnras}

\usepackage{newtxtext,newtxmath}

\usepackage[T1]{fontenc}
\usepackage{ae,aecompl}
\usepackage{graphicx}

\newcommand{\subheadstyle}[1]{{\textbf{\textit{#1}}}}

\newcommand{\Ftwo}{F^{(2)}}
\newcommand{\PBAO}{P^{\rm BAO}}
\newcommand{\bk}{{\bf k}}
\newcommand\Mpch{{\rm\; Mpc}/h}
\newcommand\hMpc{\; h/{\rm Mpc}}
\newcommand{\pref}[1]{(\ref{#1})}

\usepackage{amsmath}	
\usepackage{amssymb}	

\usepackage{layouts}

\title[BAO in the Interferometric Basis]{A Physical Picture of Bispectrum Baryon Acoustic Oscillations in the Interferometric Basis}

\author[Child et al.]{
Hillary L. Child$^{1,2}$\thanks{E-mail: childh@uchicago.edu (HC)},
Zachary Slepian$^{3,4,5}$\thanks{E-mail: zslepian@ufl.edu (ZS)} \&
Masahiro Takada$^{6}$
\\
$^{1}$HEP Division, Argonne National Laboratory, Lemont, IL 60439, USA\\
$^{2}$Department of Physics, University of Chicago, Chicago, IL 60637, USA\\
$^{3}$Department of Astronomy, University of Florida, 211 Bryant Space Sciences Center, Gainesville, FL 32611, USA\\
$^{4}$Lawrence Berkeley National Laboratory, 1 Cyclotron Road, Berkeley, CA 94720, USA\\
$^{5}$Berkeley Center for Cosmological Physics, University of California, Berkeley, Berkeley, CA 94720, USA\\
$^{6}$Kavli Institute for the Physics and Mathematics of the
Universe (WPI), UTIAS, The University of Tokyo, Chiba 277-8583, Japan}

\date{Accepted XXX. Received YYY; in original form ZZZ}

\pubyear{2018}

\begin{document}
\label{firstpage}
\pagerange{\pageref{firstpage}--\pageref{lastpage}}
\maketitle

\begin{abstract}
We present a picture of the matter bispectrum in a novel ``interferometric'' basis designed to highlight interference of the baryon acoustic oscillations (BAO) in the power spectra composing it. Triangles where constructive interference amplifies BAO provide stronger cosmic distance constraints than triangles with destructive interference. We show that the amplitude of the BAO feature in the full cyclically summed bispectrum can be decomposed into simpler contributions from single terms or pairs of terms in the perturbation theory bispectrum, and that across large swathes of our parameter space the full BAO amplitude is described well by the amplitude of BAO in a single term. The dominant term is determined largely by the $\Ftwo$ kernel of Eulerian standard perturbation theory. We present a simple physical picture of the BAO amplitude in each term; the BAO signal is strongest in triangle configurations where two wavenumbers differ by a multiple of the BAO fundamental wavelength.
\end{abstract}




\section{Introduction}

The Baryon Acoustic Oscillation (BAO) method \citep{1998ApJ...504L..57E, 2003ApJ...594..665B, 2003PhRvD..68f3004H, 2003PhRvL..90i1301L, 2003ApJ...598..720S} has become a central means of pursuing the essential nature of dark energy, a mysterious substance making up roughly $72\%$ of the present-day Universe. The BAO method uses the imprint of sound waves in the early Universe on the late-time clustering of galaxies to probe the cosmic expansion history, which through general relativity is linked to dark energy. 

The BAO method has been applied to the galaxy 2-point correlation function (2PCF) and power spectrum \citep{Ross:2016gvb, 2017MNRAS.470.2617A}, as well as more recently to the galaxy 3-point correlation function (3PCF) \citep{Gaztanaga:2008sq, Slepian:2015hca, Slepian:2016kfz} and bispectrum \citep{2018MNRAS.478.4500P}, measuring respectively 2- and 3-point clustering over random in configuration space and Fourier space. While measurements of the bispectrum and 3PCF have improved BAO constraints over those of the power spectrum alone, optimal constraints are difficult to obtain given the large number of bispectrum triangles. Bispectrum covariance matrices are often estimated from mock catalogs (\citealt{Gaztanaga:2008sq, 2018MNRAS.478.4500P}; though other approaches exist, see \citealt{Slepian:2015hca, Slepian:2016kfz}). In order to properly estimate covariance matrices, the number of mock catalogs must greatly exceed the number of triangles \citep{Percival:2013sga}; when many triangles are used to constrain BAO, the number of mocks needed is unrealistic with present computational resources. For example, \cite{2018MNRAS.478.4500P} measured the bispectrum for a large number of triangles, but noted that their error bars were limited by the number of mocks available to estimate the covariance matrix; more mocks would improve their 1.1\% precision joint constraints substantially to 0.7\%, a gain of more than 30\%. 

With better understanding of which bispectrum triangles are most sensitive to BAO, future studies could obtain better BAO constraints with a smaller set of triangles and, therefore, a smaller covariance matrix. These triangles could be identified by measuring all bispectrum triangles and their covariances, but such an approach also faces the problem of limited mock catalogs. Because fully $N$-body mocks cannot presently provide a good estimate of the full covariance matrix of all bispectrum triangles, we must select a set of optimal triangles for BAO constraints---without knowledge of the full covariance matrix. 

Recently, our work in \cite{2018arXiv180611147C} proposed one technique to select bispectrum measurements that are sensitive to BAO. We highlighted that BAO in the bispectrum constructively interfere in certain triangle configurations, amplifying the BAO signal. In a short work, we measured bispectra only on triangles where the BAO signal is amplified. With this relatively small set of bispectrum measurements we found substantial improvements in BAO constraints over power spectrum measurements alone, equivalent to lengthening BOSS by roughly 30\%. Our method for triangle selection greatly reduced the number of bispectrum measurements necessary to obtain such an improvement. 

In detail, at leading order the bispectrum involves products of two power spectra; each power spectrum introduces an oscillatory BAO feature. When the oscillations are in phase, they amplify the BAO signal. In our earlier paper, we showed that this constructive interference increases the amplitude of BAO in the bispectrum. We introduced a new parametrization of bispectrum triangles: instead of the three triangle sides $k_1$, $k_2$, and $k_3$, we use the length of one triangle side $k_1$, the difference in length of the second from the first in units of the BAO fundamental wavelength, and the angle between them $\theta$. We computed the BAO amplitude for a selection of triangle configurations, producing a map of the root-mean-square (RMS) amplitude as a function of the length difference and the opening angle. Using this RMS map, we can surgically identify the configurations most suitable for studies of BAO in the bispectrum. 

The ``interferometric basis'' proposed in our earlier work promises other applications beyond improvement in BAO constraints with relatively few bispectrum measurements. First, since our method identifies the triangles that are most sensitive to BAO, it offers an approach to more efficiently investigate the independence of bispectrum information from that obtained via reconstruction \citep{Eisensteinetal:07, 2009PhRvD..80l3501N, 2009PhRvD..79f3523P, Padmanabhanetal:12} and the covariance between the bispectrum and power spectrum. Second, our parameterization allows intuitive visualization of BAO in the bispectrum. Third, our interferometric approach is sensitive to phase shifts associated with $N_{eff}$ (such as that driven by relativistic neutrinos at high redshift \citep{Baumann:2017lmt}), spinning particles in the early Universe \citep{MoradinezhadDizgah:2018ssw}, or relative velocities between baryons and dark matter \citep{2010JCAP...11..007D, 2010PhRvD..82h3520T, 2011JCAP...07..018Y, Slepian:2016nfb}. 

In this paper, we investigate more fully the physics of BAO in the interferometric basis. The bispectrum is a sum of three cyclic terms, but for many triangles, the cyclic sum is dominated by only one or two of the three terms. Which term dominates is determined primarily by the $\Ftwo$ kernel of Eulerian standard perturbation theory (SPT). Products of power spectra also enter the leading-order perturbation theory (PT) bispectrum, but their role in the dominance structure is secondary; instead, they introduce the oscillatory features whose interference is highlighted by our basis.

When the bispectrum is dominated by only one or two terms, the amplitude of BAO in the full bispectrum can be approximated by the BAO amplitude in the dominant term or terms. We show analytically that because BAO are a small feature in the power spectrum, the difference between the RMS amplitude computed under this approximation and the full RMS amplitude vanishes at leading order. We can therefore decompose the full RMS map into approximate ``eigen-RMSes'' computed from individual terms. Numerical work verifies that this decomposition successfully reproduces the primary features of the full RMS map.

To understand these features, we study the behavior of BAO in each term and pair of terms that can dominate the bispectrum. The structure of the power spectrum, in particular BAO and their envelope due to Silk damping, determines the RMS amplitude. For each triangle configuration, the BAO amplitude in each term is driven by one of four interactions between power spectra: interference, incoherence, feathering, or single power spectrum. The first, interference, can dramatically amplify BAO amplitude in certain configurations, like those used in \cite{2018arXiv180611147C} to constrain the BAO scale.

In general, then, the $\Ftwo$ kernel determines which pairs of power spectra set the amplitude of BAO in the measured bispectrum. The remainder of the paper details this broad picture as follows. In \S\ref{sec:interferometric_basis}, we review the interferometric basis as presented in our earlier work, and define the RMS amplitude of the ratio of physical to ``no-wiggle'' bispectrum we use in our analysis. \S\ref{sec:notation} introduces notation used throughout. \S\ref{sec:dominance_regions} shows which triangle configurations are dominated by a single term of the bispectrum cyclic sum; in these regions, the BAO signal in the bispectrum simplifies to the BAO signal in a single term, as shown analytically in \S\ref{sec:analytic_eigen}. In \S\ref{sec:eigenvariances}, we numerically calculate the BAO amplitude in each term of the cyclic sum. These ``eigen-RMSes'' approximate the BAO amplitude in the full bispectrum in the regions where the corresponding terms dominate, and they assemble into a picture that matches the full map of BAO amplitude. \S\ref{sec:discussion} presents implications of our study for the reduced bispectrum, the 3PCF, and multipole expansions of the 3PCF and bispectrum. \S\ref{sec:conclusion} concludes.

Throughout we adopt a spatially flat $\Lambda$CDM cosmology at $z=0$ consistent with the WMAP-7 \citep{2011ApJS..192...18K} parameters of the MockBOSS simulations \citep{Sunayamaa:2015aba} used in \cite{2018arXiv180611147C}: ${\rm \Omega}_{\rm m} = 0.2648$, ${\rm \Omega}_{\rm b}h^2 = 0.02258$, $n_s = 0.963$, $\sigma_8 = 0.80$, and $h=0.71$.

\section{Interferometric Basis}
\label{sec:interferometric_basis}
Several sets of triangle parameters have been used in previous works for the isotropic bispectrum and three-point correlation function (3PCF). \cite{2010MNRAS.406.1014S}, \cite{2013MNRAS.429..344K}, and \cite{Baldauf:2014qfa} considered the bispectrum for equilateral or isosceles triangles. \cite{Scoccimarro:1997st}, \cite{2000ApJ...544..597S}, \cite{2002PhR...367....1B}, \cite{2010MNRAS.406.1014S}, \cite{2012PhRvD..86h3540B}, \cite{2012JCAP...02..047G}, and \cite{2017PhRvD..96d3513H} used one side $k_1$, the ratio of a second side to the first $k_2/k_1$, and the angle between the two $\theta_{12}$; the third side can also be parameterized simply as $k_3$ \citep{2018MNRAS.478.4500P} or by its ratio to $k_1$ \citep{2009ApJ...703.1230J, Baldauf:2014qfa}. \cite{2000ApJ...544..597S} and \cite{2006PhRvD..74b3522S} allowed each triangle side to be any integer multiple of the bin width $\Delta k \simeq 0.015 \hMpc$.

Like the bispectrum, the 3PCF can be parametrized using one side $r_1$, the ratio of a second side to the first $r_2/r_1$, and the opening angle $\theta$ \citep{Kulkarni:2007qu, 2011ApJ...737...97M, Marin:2013bbb}. Other parameterizations use two sides $r_1$ and $r_2$. The third parameter can be the opening angle $\theta$ between them \citep{2011ApJ...726...13M, 2011ApJ...739...85M, 2014ApJ...780..139G, 2015MNRAS.449L..95G}, the cosine $\mu$ of the opening angle \citep{Gaztanaga:2008sq}, a shape parameter combining the three side lengths \citep{1980lssu.book.....P, 2004ApJ...607..140J, 2004MNRAS.353..287W, Nichol:2006mg}, or a multipole expansion of the dependence on opening angle \citep{2004ApJ...605L..89S, Pan:2005ym, Slepian:2015hca, Slepian:2016kfz}. Many studies of the anisotropic bispectrum and 3PCF, which retain information about the line of sight, have also employed a multipole basis with respect to the line of sight in redshift space (e.g. \citealt{2017MNRAS.467..928G, 2017PhRvD..95d3528Y, 2018MNRAS.476.4403C, 2018arXiv180604015D, 2018JCAP...07..038N, 2018arXiv180707076Y}). \cite{Slepian:2017lpm} and \cite{2018arXiv180302132S} use a spherical harmonic expansion of the 3PCF and bispectrum, which includes information on both the internal angle and the angles to the line of sight.

In \cite{2018arXiv180611147C}, we introduced a new basis for bispectrum work motivated by the physics of BAO in the bispectrum. The bispectrum $B$ involves products of linear matter power spectra $P$ (e.g. \citealt{Scoccimarro:1997st}), as
\begin{equation}
\label{eqn:B0}
B(k_1, k_2, k_3) = 2 P(k_1)P(k_2)\Ftwo(k_1, k_2; \hat{\bk}_1\cdot\hat{\bk}_2) + \mbox{cyc.} 
\end{equation}
We refer to $2P(k_1)P(k_2)\Ftwo(k_1, k_2; \hat{\bk}_1\cdot\hat{\bk}_2)$ as the pre-cyclic term, and to the terms denoted by cyc. as the post-cyclic terms. $\Ftwo$ is the Eulerian standard perturbation theory kernel that generates a second-order density field when integrated against two linear density fields. The $\Ftwo$ kernel depends only on two side lengths $k_i$ and $k_j$ and the angle between them (through the dot product $\hat{\bk}_i\cdot\hat{\bk}_j$):
\begin{multline}
\label{eqn:F2}
\Ftwo(k_i, k_j; \hat{\bk}_i\cdot\hat{\bk}_j)\\
=\frac{5}{7}+\frac{1}{2}\left(\frac{k_i}{k_j}+\frac{k_j}{k_i}\right)(\hat{\bk}_i\cdot\hat{\bk}_j)+\frac{2}{7}(\hat{\bk}_i\cdot\hat{\bk}_j)^2.
\end{multline}
We will refer to the middle term of equation~\pref{eqn:F2}, $(k_i/k_j+k_j/k_i)$, as the dipole contribution to $\Ftwo$.

\begin{figure}
    \centering
    \includegraphics[width=0.7\linewidth]{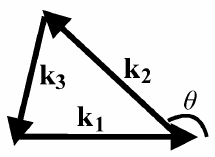}
    \caption{The triangle parameter $\theta$ is defined in equation \pref{eqn:delta_theta_def} as the exterior angle between $\bk_1$ and $\bk_2$.}
    \label{fig:triangle_theta_def}
\end{figure}

The power spectra in equation~\pref{eqn:B0} have BAO features that oscillate sinusoidally. These features can thus interfere, motivating us to consider a parameterization of the bispectrum that transparently captures the phase structure. In particular, we set up our parametrization to capture the phase structure of the pre-cyclic term (first term in equation \ref{eqn:B0}) in $k_1$ and $k_2$, as follows:
\begin{align}
\label{eqn:delta_theta_def}
k_1,\;\;\; k_2 - k_1 = \delta \left( \frac{\lambda_f}{2}\right), \;\;\;\cos \theta = \hat{\bk}_1\cdot\hat{\bk}_2.
\end{align}
Throughout this work we assume without loss of generality that $\delta$ is positive, so $k_2 > k_1$. The external angle $\theta$ is shown in Figure \ref{fig:triangle_theta_def}. The fundamental wavelength of the BAO in Fourier space $\lambda_f$ is given by
\begin{align}
\label{eqn:def_lambda_f}
\lambda_f = \frac{2\pi}{\tilde{s}_f} \approx 0.0574 \hMpc,
\end{align}
where $\tilde{s}_f = 109.5\Mpch$ is the effective sound horizon evaluated at a fiducial wavenumber $k_f = 0.2\hMpc$. As discussed in \cite{EisensteinHu:98}, the lowest-wavenumber nodes of the baryonic transfer function occur at higher wavenumber than the nodes of $\sin{ks}$. The effective sound horizon grows with $k$ for $k \lesssim 0.05\hMpc$ and asymptotes to the sound horizon $s$ for $k \gtrsim 0.05\hMpc$; we define $\lambda_f$ according to its asymptotic value at $k_f = 0.2\hMpc$.

We use ``configuration'' to describe a set of triangles with fixed $\delta$ and $\theta$ over the range $0.01 \leq k_1/[h/{\rm Mpc}] \leq 0.2$. For each configuration, we divide $k_1$ into 100 bins of width $1.9 \times 10^{-3} \hMpc$. The other two wavenumbers $k_2$ and $k_3$ are calculated from each $k_1$ according to equation~\pref{eqn:delta_theta_def}. We study configurations with $0 \leq \delta \leq 4.25$ and $0 \leq \theta \leq 1.1$, with 80 points in $\theta$ and 80 in $\delta$ for a total of 6400 configurations. We sample many $(\delta, \theta)$ points simply to produce well-resolved figures, but we note that configurations that are close to each other in $(\delta, \theta)$ space are highly covariant. 

We restrict $k_1$ to the range $0.01 \leq k_1/[h/{\rm Mpc}] \leq 0.2$ and $\delta$ to be less than about 4 to capture most of the effects of BAO. For most configurations, $k_1$ is smaller than the other two triangle sides, but for small $\delta$ and $\theta/\pi \sim 1$, $k_3$ can be smaller than $k_1$. In our previous paper, these configurations were not used to constrain BAO because they are subject to cosmic variance and covariant with the power spectrum (as discussed under ``Simulations'' in \citealt{2018arXiv180611147C}).

Below our minimum wavenumber of $0.01\hMpc$, cosmic variance becomes significant; given the mock catalogs we use one cannot make a sufficient number of subdivisions to estimate covariances on these large scales. Of course at large scales the covariance should be dominated by the Gaussian Random Field (GRF) contribution, so a template could be used to model the covariance (e.g., as \citealt{Slepian:2015qza} does for the isotropic 3PCF and \citealt{Slepian:2017lpm} for the anisotropic 3PCF). 

However, even with an adequate covariance, the contribution of low-wavenumber modes to BAO constraints should be small given the small number of large-scale modes in the volume of a survey such as DESI\footnote{\url{http://desi.lbl.gov}} \citep{2016arXiv161100036D}. Our minimum wavenumber corresponds to a physical scale of $628 \Mpch$; DESI will have volume of order $50 \; [{\rm Gpc}/h]^3$ equivalent to a box side length of roughly $3700 \Mpch$. Thus there are of order 200 modes of wavelength $628 \Mpch$ in the box, enabling measurement to about 7\% precision. The contribution of low-wavenumber bispectrum modes to BAO constraints will therefore be negligible compared to the 0.1\% precision DESI will achieve using power spectrum BAO at higher wavenumber. It \emph{is} the case that the other two triangle sides can probe higher wavenumbers---our maximum $\delta$ studied is $4$, so $k_1 = 0.01 \hMpc$ corresponds to at most $k_2 = 2 \lambda_f + 0.01 \hMpc = 0.125 \hMpc$ and $k_3 = k_1+k_2 = 0.135 \hMpc$. These wavenumbers do access BAO scales, but nonetheless, the bispectrum error bar will not be competitive with DESI power spectrum precision as the total bispectrum error bars of such configurations will be dominated by the cosmic variance of the shortest side $k_1$. 

At higher wavenumbers than our maximum, even at the level of linear theory Silk damping \citep{1968ApJ...151..459S} degrades the BAO signal. The Silk damping scale $k_{\rm Silk}$ (equation~7 of \citealt{EisensteinHu:98}), is approximately $0.125 \hMpc$ for our cosmology; at wavenumbers above $k_{\rm Silk}$, the BAO signal in the transfer function is increasingly suppressed as ${\rm exp}[-(k/k_{\rm Silk})^{1.4}]$ (equation~21 of \citealt{EisensteinHu:98}). 

Wavenumbers above $k_{\rm NL} \sim 0.1 \hMpc$ are nonlinear, so perturbation theory no longer provides an accurate model of the bispectrum at these scales \citep{2012JCAP...06..018R}. Effective field theory (EFT) models perform reasonably well up to $k \sim 0.3 \hMpc$; \cite{Carrasco:2012cv} describes the power spectrum to the percent level for $k \lesssim 0.3 \hMpc$. In the case of the bispectrum (e.g. \citealt{Bertolini:2016hxg, Nadler:2017qto, 2018arXiv180406849D}), the maximum wavenumber at which EFT agrees with simulations depends on configuration and cosmology, but EFT models of the real-space matter bispectrum perform well up to $k \sim 0.2 \hMpc$ \citep{Angulo:2014tfa, Baldauf:2014qfa}. In redshift space, however, perturbation theory models break down at yet smaller wavenumbers, differing from bispectrum measurements at the 10\% level by $k=0.1 \hMpc$ \citep{Smith:2007sb}. 

Baryonic effects, which are not as yet satisfactorily modeled, also become important at wavenumbers above our maximum. For $k \gtrsim 1 \hMpc$, hydrodynamical simulations find a $5-15\%$ alteration in the power spectrum relative to dark-matter-only simulations \citep{2018MNRAS.480.3962C}. As the bispectrum scales roughly as $P^2$ with $P$ the power spectrum, this $\sim$$10\%$ uncertainty in the power spectrum likely translates to $\sim$$20\%$ in the bispectum, which at tree level is proportional to the square of the power spectrum (see also the hierarchical ansatz of \citealt{1977ApJ...217..385G}). In the absence of a theoretical model for baryonic effects, the uncertainty in high-wavenumber models of the bispectrum is much too large to measure BAO to sub-percent precision. 

\begin{figure}
\includegraphics{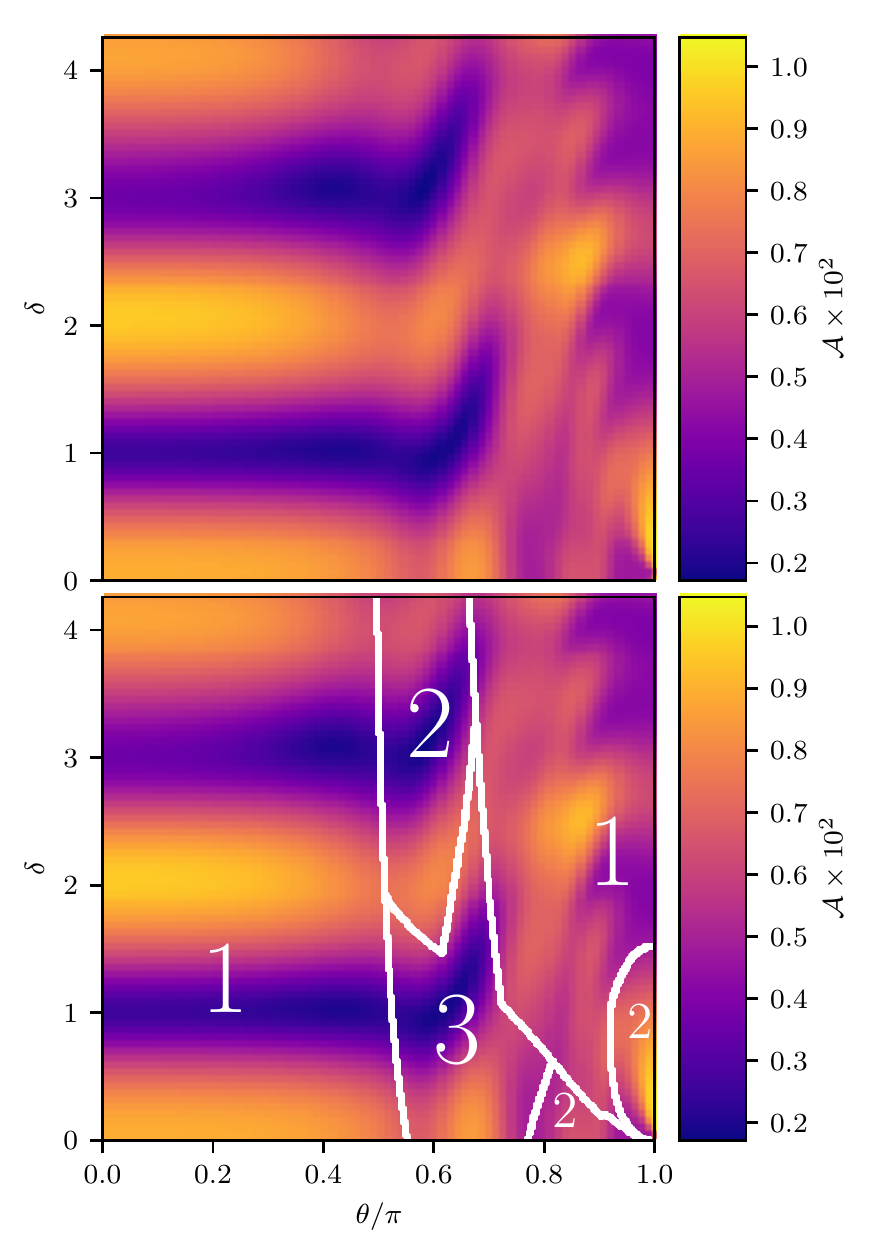}
\caption{\label{fig:full_colorplot} \emph{Top---}The root-mean-square amplitude $\mathcal{A}$ (equation \ref{eqn:mathcalA}) of the bispectrum BAO feature in triangle configurations parameterized by $(\delta, \theta)$. Maxima and minima are set by the constructive and destructive interference of BAO oscillations in the bispectrum. \emph{Bottom---}In many regions of the RMS map, a single term or pair of terms in the bispectrum cyclic sum \pref{eqn:B0} dominates the sum. The boundaries between these regions (white lines) correspond to changes in the behavior of the RMS map. The numerals indicate the number of terms that must be considered to accurately approximate the bispectrum.}
\end{figure}

Overall, then, the range of wavenumbers we consider is a conservative cut to isolate the regime where BAO effects are most prominent and the bispectrum is best understood. Within this range of scales, our interferometric basis identifies the configurations where constructive interference of power spectra amplifies the BAO ``wiggles.'' To quantify the presence of BAO in each configuration, we compute the RMS amplitude of the ratio $R$ of the bispectrum $B(k_1,\delta,\theta)$ to its no-wiggle analog $B^{\rm nw}(k_1,\delta,\theta)$. We have
\begin{equation}
\label{eqn:R_bispec}
R(k_1,\delta,\theta) = \frac{B(k_1,\delta,\theta)}{B^{\rm nw}(k_1,\delta,\theta)},
\end{equation}
where the numerator is computed using power spectra $P(k)$ from CAMB \citep{Lewis:1999bs} and the denominator using power spectra $P^{\rm nw}(k)$ from the fitting formula for the no-wiggle transfer function of \cite{EisensteinHu:98}. The variance is
\begin{equation}
\label{eqn:mathcalA}
\mathcal{A}^2(\delta,\theta) \equiv \int_{0.01}^{0.2} \left[R(k_1,\delta,\theta) - \bar{R}(\delta,\theta) \right]^2 
\frac{\mathrm{d}k_1}{[h/{\rm Mpc}]},
\end{equation}
where $\bar{R}(\delta, \theta)$ is the mean of $R(k_1, \delta, \theta)$ on the same range, $0.01 \leq k_1/[h/{\rm Mpc}] \leq 0.2$. Figure~\ref{fig:full_colorplot} shows the root-mean-square amplitude $\mathcal{A}$ for a selection of configurations. Throughout this work we will refer to Figure~\ref{fig:full_colorplot} as the root-mean-square (RMS) map. 

Our basis is a transformation of the triangle sides $(k_1, k_2, k_3)$; the axes $\theta$ and $\delta$ of our RMS map correspond roughly to $k_3$ and $k_2$, respectively. In our basis, the wavenumbers $k_2$ and $k_3$ depend on $k_1$, $\delta$, and $\theta$ as
\begin{equation}
\begin{split}
\label{eqn:k2_k3_fcn_delta_theta}
k_2 &= k_1 + \delta \lambda_f / 2, \\
k_3 &= \sqrt{k_1^2 + k_2^2 + 2 k_1k_2 \cos{\theta}}\\
&= \sqrt{k_1 \left(1+\cos{\theta}\right)\left( 2k_1 + \delta \lambda_f \right) + \left(\delta \lambda_f/2 \right)^2},
\end{split}
\end{equation}
where the first equality for $k_3$ stems from the orientation of $\theta$ shown in Figure~\ref{fig:triangle_theta_def} and the law of cosines. Configurations with the same $k_3$ therefore lie along sloped curves in the $(\delta, \theta)$ plane. As the power spectrum depends only on the magnitude of the wavenumber, these curves are also traces of constant $P(k_3)$.

\begin{figure}
    \centering
    \includegraphics{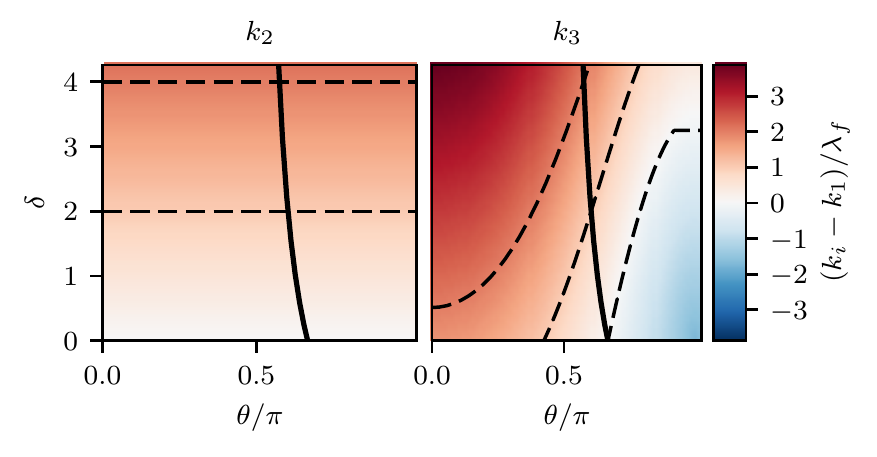}
    \caption{In our basis, $k_2$ depends only on $\delta$, while $k_3$ varies with both $\delta$ and $\theta$. The behavior of these two wavenumbers in the $(\delta, \theta)$ basis is critical for understanding both the power spectrum and $\Ftwo$ kernel. As $P(k) \sim k$, the structure of $P(k)$ in the $\delta$-$\theta$ plane is similar to that of the individual wavenumbers, while $\Ftwo$ is a more complicated function (as shown in Figure~\ref{fig:f2_kernel_detail}). $k_2$ and $k_3$ are calculated according to equation~\pref{eqn:k2_k3_fcn_delta_theta} with $k_1 = 0.1 \hMpc$. Dashed lines in the left panel show configurations for which $k_2 = k_1 + n\lambda_f$ (equation \ref{eqn:k1_k2_equal_delta}) for $n=1$ and $2$; $n=0$ coincides with the $\theta$ axis. In the right panel, dashed curves show configurations for which $k_3 = k_1 + n\lambda_f$ (equation \ref{eqn:k1_k3_equal_delta}) for $n=0,1$, and $2$. Solid curves show configurations for which $k_2=k_3$; the color is red where $k_i$ is larger than $k_1$, blue where $k_i$ is smaller than $k_1$, and white where $k_i=k_1$.  }
    \label{fig:k2_k3_magnitude}
\end{figure}

The parameter $\delta$ was chosen to produce constructive interference in the precyclic term of the bispectrum (first term in equation \ref{eqn:B0}). However, constructive interference is not limited only to this single term: we expect interference as well where $k_2$ and $k_3$, or $k_3$ and $k_1$, differ by integer multiples $n$ of the BAO wavelength $\lambda_f$. We calculate the configurations for which these conditions are satisfied. Curves where $k_2 = k_1 + n \lambda_f$ are horizontal lines in the $(\theta, \delta)$ plane, as shown in the left panel of Figure~\ref{fig:k2_k3_magnitude}; that is, $k_2 = k_1 + n \lambda_f$ where
\begin{equation}
\label{eqn:k1_k2_equal_delta}
    \delta/2 = n.
\end{equation}
The curves where $k_2 = k_3 + n \lambda_f$, shown for $n=0$ in Figure~\ref{fig:k2_k3_magnitude}, are given by
\begin{equation}
\label{eqn:k2_k3_equal_delta}
\delta = \frac{-2k_1^2\cos{\theta} - k_1^2 - 2 n \lambda_f k_1 + n^2 \lambda_f^2}{n \lambda_f^2 +\lambda_f k_1 \cos{\theta}}.
\end{equation}
We only show the $n=0$ case as higher harmonics of $k_2 = k_3 + n \lambda_f$ do not correspond to features in the RMS map, as discussed in \S\ref{subsubsec:eigenvariance_incoherent} below. The curves where $k_3=k_1 + n \lambda_f$, shown as dashed curves in the right panel of Figure~\ref{fig:k2_k3_magnitude}, follow
\begin{equation}
\label{eqn:k1_k3_equal_delta}
\delta = \frac{2}{\lambda_f}\left[\sqrt{k_1 \left(k_1 \cos^2\theta + n \lambda_f \right) + n^2 \lambda_f^2} - k_1 (1+\cos \theta) \right].
\end{equation}

In general, equations~\pref{eqn:k2_k3_equal_delta} and \pref{eqn:k1_k3_equal_delta} depend on both $k_1$ and $\theta$. In the special case of the equilateral configuration, the $k_1$ dependence cancels; that is, when $\theta/\pi=2/3=0.67$ and $n=0$, equation \pref{eqn:k2_k3_equal_delta} gives $\delta=0$. For these configurations, $k_3$ equals $k_2$ for all $k_1$. For all other configurations, however, equations~\pref{eqn:k2_k3_equal_delta} and \pref{eqn:k1_k3_equal_delta} can only be satisfied for a single $k_1$. When necessary, we choose a representative value of $k_1=0.1 \hMpc$ to compute the configurations for which equations~\pref{eqn:k2_k3_equal_delta} and \pref{eqn:k1_k3_equal_delta} hold.

\section{Notation}
\label{sec:notation}
Here we define notation for several combinations of power spectra, $\Ftwo$ kernels, and bispectra that will be used throughout.

BAO in the bispectrum come only from oscillations in the power spectrum, which we isolate as 
\begin{equation}
    \PBAO_i = \frac{P(k_i)}{P^{\rm nw}(k_i)} \equiv 1 + w_i,
\label{eqn:PBAO_wi_def}
\end{equation}
where $w_i$ is defined through this equality and represents the BAO-only piece of the power spectrum. We note that $w_{i} \ll 1$; the baryon fraction $f_{\rm b}$ in our Universe is small ($f_{\rm b} \equiv \Omega_{\rm b}/\Omega_{\rm m} \sim 20\%$), so the BAO are a small feature in the power spectrum.

Each term in the cyclic sum \pref{eqn:B0} is denoted by
\begin{equation}
    B_{ij} = 2P_{ij}\Ftwo_{ij}
\end{equation}
with 
\begin{equation}
\label{eqn:pij_def}
P_{ij} = P(k_i)P(k_j)
\end{equation}
and
\begin{equation}
\Ftwo_{ij} = \Ftwo(k_i, k_j; \hat{\bk}_i\cdot\hat{\bk}_j).
\end{equation}
The ratio of each term to its no-wiggle analog is
\begin{align}
\label{eqn:Rij_def}
    R_{ij} = \frac{B_{ij}}{B_{ij}^{\rm nw}} 
    = \frac{P_{ij}}{P_{ij}^{\rm nw}} 
    = \PBAO_i\PBAO_j,
\end{align}
where the second equality holds because the $\Ftwo$ kernel is unaltered going from a physical to a ``no-wiggle'' cosmological model. The kernel stems from the Newtonian gravity solution of the equations of perturbation theory assuming an Einstein-De Sitter (matter-dominated) cosmology, and is thus independent of the input linear power spectrum. We note that all cosmological parameters of the no-wiggle model, including the matter density, are identical to the physical model.

Using the definition \pref{eqn:PBAO_wi_def} of $w$, we may rewrite 
\begin{align}
R_{ij} = 1 + w_i + w_j + w_i w_j \equiv 1 + w_{ij}    
\label{eqn:Rij_wij_def}
\end{align}
where the last equality defines $w_{ij}$, the oscillatory piece of one term of the bispectrum \pref{eqn:B0}. 

To refer to a ratio where one term in the sum is negligible, we use
\begin{equation}
    \label{eqn:R_ijjkdef}
    R_{ij+jk} = \frac{B_{ij}+B_{jk}}{B_{ij}^{\rm nw}+B_{jk}^{\rm nw}}.
\end{equation}
The sum $R_{12}+R_{23}+R_{31}$ is not equal to $R$ of equation~\pref{eqn:R_bispec}.

\section{Regions of Dominance}
\label{sec:dominance_regions}
In order to understand the behavior of the BAO amplitude shown in the RMS map (Figure~\ref{fig:full_colorplot}), we seek to identify configurations where the cyclic sum of the perturbation theory bispectrum \pref{eqn:B0} simplifies. That is, we ask whether there are any regions where the behavior of the full bispectrum is determined by only one or two of the three terms in the cyclic sum. 

\begin{figure}
    \centering
    \includegraphics{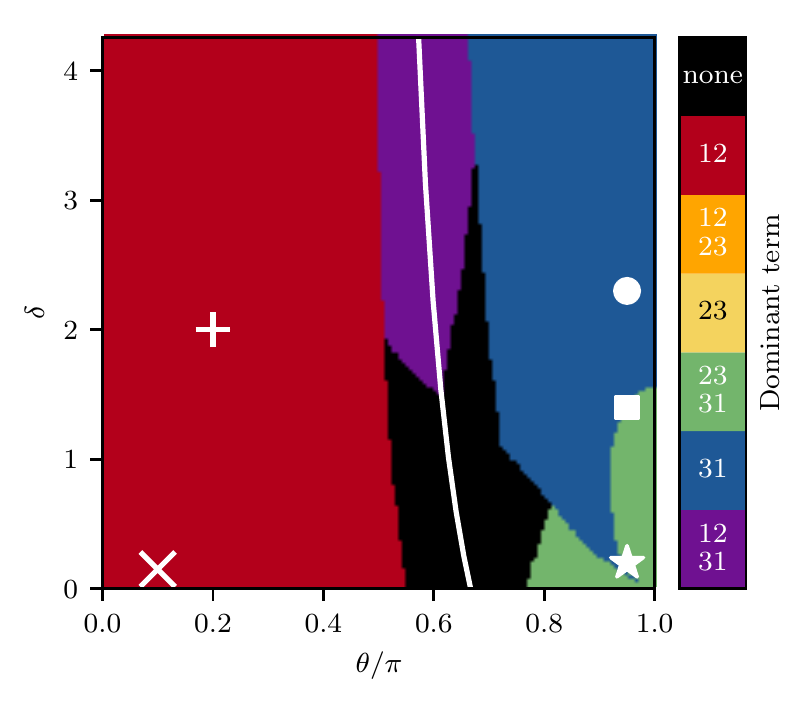}
    \caption{The dominance map (\S\ref{sec:dominance_regions}) shows regions where the bispectrum cyclic sum simplifies to a single term or pair of terms. A term dominates (\S\ref{subsec:definition_of_dominance}) if the median of its ratio with each of the other two terms is at least $5$. In our color scheme, primary colors (red, yellow, and blue) represent single terms, while the secondary colors (orange, green, and purple) represent pairs of terms. For an equilateral configuration ($\delta=0$, $\theta/\pi=0.67$), all three terms are identical so none can dominate; the black region surrounds this configuration. Symbols indicate representative configurations that are discussed in detail in \S\ref{subsec:regions_of_the_dominance_map}. The white curve shows $k_2=k_1$ (equation \ref{eqn:k2_k3_equal_delta}), where at least two terms must be of comparable magnitude (\S\ref{subsubsec:dominance_black_purple}). }
    \label{fig:dominance_plot_guide}
\end{figure}
Figure~\ref{fig:dominance_plot_guide}, our ``dominance map,'' shows that many of the configurations are indeed dominated by a single term (red and blue regions), and others are dominated by two terms while the third is negligible (green and purple regions). The RMS map (Figure~\ref{fig:full_colorplot}) reflects the dominance structure. The horizontal bands at $\theta/\pi \lesssim 0.4$ (red region, $B_{12}$ dominant) transition to sloped bands in the purple and black regions. The blue region ($B_{31}$ dominant) corresponds to a pattern of small maxima and minima in the RMS map that we call ``feathering.'' Finally, in the green region at low $\delta$ and high $\theta$, RMS amplitude is maximized for triangles where two wavevectors are nearly antiparallel and the third is small. The mechanisms that drive these different patterns in each region are described in detail in \S\ref{sec:eigenvariances} below. 

In this section, we detail the calculation and behavior of the dominance map (Figure~\ref{fig:dominance_plot_guide}). In \S\ref{subsec:definition_of_dominance}, we present our definitions of dominance, which require the choice of a factor $f$. The specific ways in which triangle geometry determines which term dominates are discussed in detail in \S\ref{subsec:regions_of_the_dominance_map}; the dominance map is driven primarily by the behavior of the $\Ftwo$ kernel reinforced by the broadband behavior of the power spectrum, as we will further detail in \S\ref{subsec:f2_kernel_vs_pk_dominance}. In most regions of the dominance map, the maximum and minimum terms in the $\Ftwo$ kernel also maximize or minimize the power spectrum; the exceptions are discussed in \S\ref{subsec:ordering_of_subdominant_terms}. In general, in the squeezed limit where one side of the triangle can be much larger than the smallest, terms including the largest wavenumber are small. In the other limit, an equilateral triangle, all three sides are similar so all functions of them are similar as well and no sides dominate. The dominance plot shows the transitions between these two regimes. 

We note that we assume positive $\delta$, and for $\delta>0$, no region is dominated by $B_{23}$ (yellow) or the pair of terms $B_{12}+B_{23}$ (orange). When $\delta <0$, $k_1$ and $k_2$ interchange. This would correspond to mirroring across the $\theta$-axis; blue would become yellow.

\subsection{Definition of ``Dominance'' and Choice of Dominance Ratio \texorpdfstring{$f$}{f}}
\label{subsec:definition_of_dominance}
We identify dominant terms by comparing the magnitudes of terms $B_{ij}^{\rm nw}$ across $k_1$. The dominance structure is determined by the broadband behavior of the bispectrum terms, so we use the no-wiggle bispectrum $B_{ij}^{\rm nw}$ to fully isolate the broadband. Results are similar when the full bispectrum  $B_{ij}$ is used instead, as BAO are small relative to the broadband.

At each ($\delta$, $\theta$) configuration, we calculate the ratios between each pair of terms as a function of $k_1$. We then compare the medians, denoted ${\rm med}$, of these ratios to a factor $f$. We use the median because it is a smooth function of our parameters $\delta$ and $\theta$, unlike the mean, which can be skewed by large ratios between the terms at small $k_1$. The median is more representative of the typical ratio across all $k_1$ we consider. 

\subheadstyle{Dominance criterion---}If $B_{ij}$ exceeds each other term
by at least a factor of $f$, that is, if
\begin{equation}
\label{eqn:single_dominance_criterion}
    {\rm med}\left\lvert\frac{B_{ij}^{\rm nw}}{B_{jk}^{\rm nw}}\right\rvert > f,  \,\,\, {\rm med}\left\lvert\frac{B_{ij}^{\rm nw}}{B_{ik}^{\rm nw}}\right\rvert > f,
\end{equation}
we consider $B_{ij}$ dominant.

\subheadstyle{Double dominance criterion---}If two of the terms that enter the bispectrum determine its behavior while the third is relatively small, we say that two terms are double dominant. Two ratios must be within a factor of $f$ of each other but both exceed the third by at least a factor of $f$. Because terms can be either positive or negative, this comparison alone is not obviously sufficient; one term could be large and positive, and the other large and negative, such that their sum is smaller than the third term. Therefore we also require the sum of the two dominant terms to exceed the third term by a factor $f$. Our double dominance criterion is thus a set of four conditions: $B_{ij}$ and $B_{ik}$ are both dominant and only $B_{jk}$ is negligible when
\begin{equation}
\label{eqn:double_dominance_criterion}
\begin{split}
    {\rm med}\left\lvert\frac{B_{ij}^{\rm nw}}{B_{jk}^{\rm nw}}\right\rvert > f, \,\,\, {\rm med}\left\lvert\frac{B_{ik}^{\rm nw}}{B_{jk}^{\rm nw}}\right\rvert > f, \,\,\, 
    {\rm med}\left\lvert\frac{B_{ij}^{\rm nw}}{B_{ik}^{\rm nw}}\right\rvert < f, \,\,\,
    \\
    {\rm med}\left\lvert\frac{B_{ij}^{\rm nw}+B_{ik}^{\rm nw}}{B_{jk}^{\rm nw}}\right\rvert > f.
\end{split}
\end{equation}
In practice, the final condition is not relevant for any configuration we test; the differences between large positive and negative terms remain much larger than the third term, for example in the region described in further detail in \S\ref{subsubsec:dominance_green}.

 \subheadstyle{No term dominant---}If the medians of all three ratios are within $f$ of each other, then no term is dominant. 
 
\begin{figure}
\includegraphics{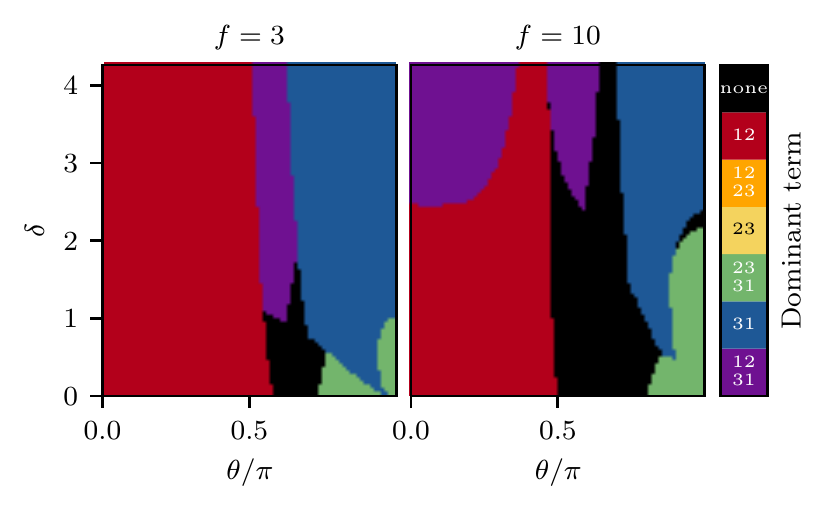}
\caption{\label{fig:dominant_terms} The shapes and locations of regions of dominance are not highly sensitive to $f$, the factor by which a term must exceed all others to be called ``dominant" (see \S\ref{subsec:definition_of_dominance}, equations \ref{eqn:single_dominance_criterion} and \ref{eqn:double_dominance_criterion}). As $f$ increases from the left panel to the right, the black and purple regions (where no single term exceeds all others by at least a factor of $f$) expand. That is, when the criterion for a single term to dominate is more strict, fewer configurations are dominated by a single term.}
\end{figure}

The dominance region plot is weakly dependent on the choice of the factor $f$, as shown in Figure~\ref{fig:dominant_terms}. As the threshold for dominance rises, less of the plane is dominated by a single term; the $B_{12}+B_{31}$-dominant, $B_{23}+B_{31}$-dominant, and no-term-dominant regions encroach on the single-term-dominant regions. We choose $f=5$ as our standard threshold for dominance, as it is sufficiently large to separate the term that dominates the RMS amplitude. With this choice of $f$, the non-dominant terms are typically less than 20\% of the dominant term, so the ratio of the bispectrum to the no-wiggle bispectrum~\pref{eqn:R_bispec} can be Taylor-expanded about the ratio $R_{ij}$ of a single term in the cyclic sum to its no-wiggle analog (as we do in \S\ref{sec:analytic_eigen}). 

\subsection{\texorpdfstring{$\Ftwo$}{F2} Kernel Drives Dominance Structure}
\label{subsec:f2_kernel_vs_pk_dominance}

\begin{figure*}
\includegraphics{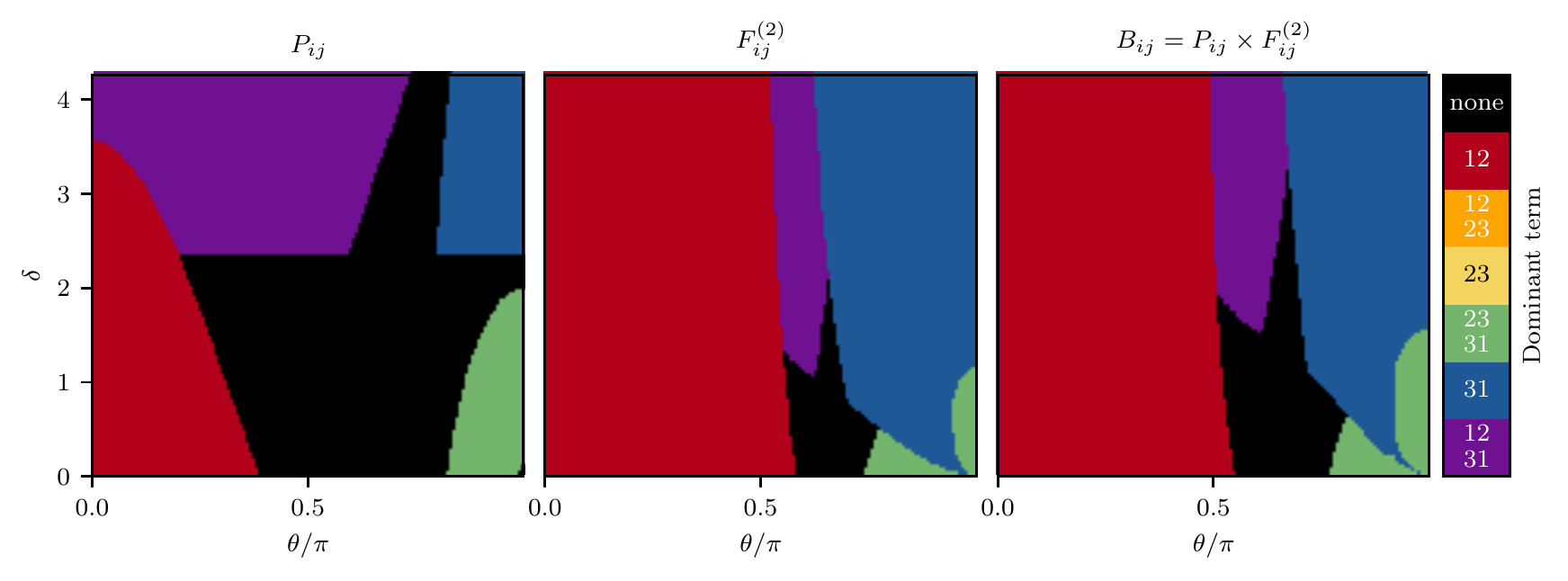}
\caption{\label{fig:dominant_terms_p_f2} The regions of dominance are determined primarily by the $\Ftwo$ kernel, as discussed in \S\ref{subsec:f2_kernel_vs_pk_dominance}; the structure of the full dominance plot (right panel) is very similar to that of the $\Ftwo_{ij}$ dominance plot (middle panel), with some modification from the products of power spectra $P_{ij}$ (left panel). For the left and middle panels, $P_{ij}$ and $\Ftwo_{ij}$, a term is dominant if exceeds the other two by a factor of $\sqrt{5}$ (chosen for consistency with $f=5$ for the product $P_{ij}\Ftwo_{ij}$). For the third panel, a term is dominant if the median of its ratio with each of the other terms is at least 5 (as in Figure~\ref{fig:dominance_plot_guide}).  }
\end{figure*}

\begin{figure*}
    \centering
    \includegraphics{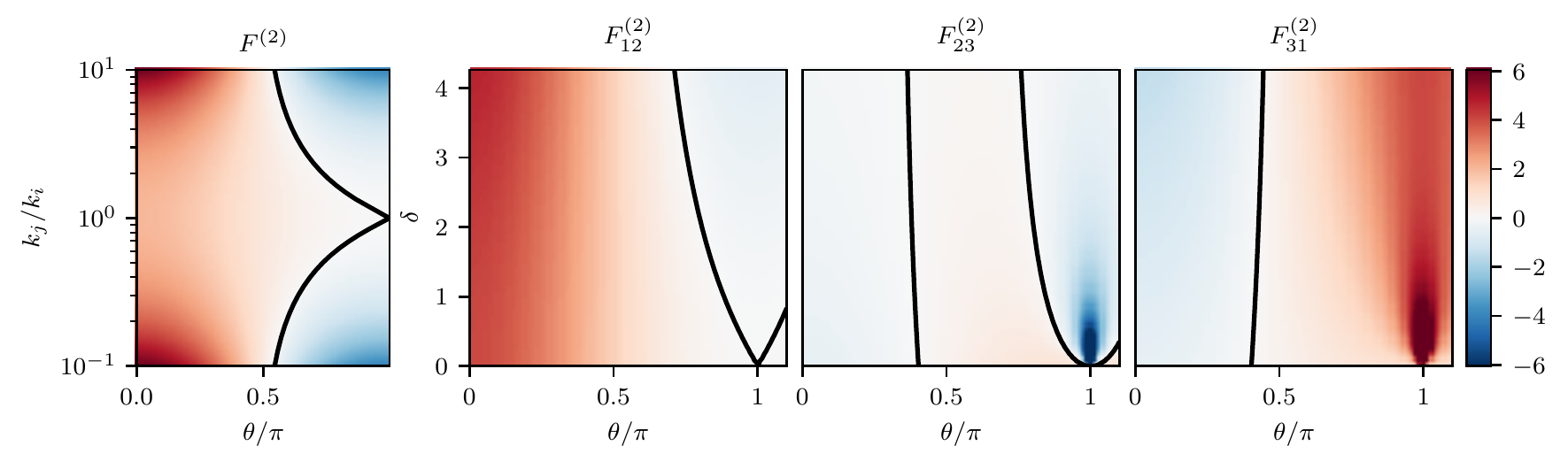}
    \caption{The $\Ftwo$ kernel drives the structure of the dominance plot. As shown in the left panel, $\Ftwo$ depends only on the angle between two sides through $\theta$ (Figure~\ref{fig:triangle_theta_def}) and the ratio of their lengths $k_j/k_i$. The kernel can be positive or negative, and crosses zero (black curves). The dynamic range therefore exceeds that of the power spectrum product $P_{ij}$, which varies only by a factor of $500$ across the triangles shown. The remaining three panels show the $\Ftwo_{ij}$ that enter the bispectrum, evaluated at $k_1 = 0.1 \hMpc$. These three panels determine the behavior of the $\Ftwo$ dominance plot (middle panel of Figure~\ref{fig:dominant_terms_p_f2}): $\Ftwo_{12}$ (middle left panel) is the largest $\Ftwo_{ij}$ in the red region of the middle panel of Figure~\ref{fig:dominant_terms_p_f2}, $\Ftwo_{31}$ (right panel) dominates in the blue region of Figure~\ref{fig:dominant_terms_p_f2}, and in the green region of Figure~\ref{fig:dominant_terms_p_f2}, both $\Ftwo_{23}$ (middle right panel) and $\Ftwo_{31}$ are large while $\Ftwo_{12}$ is small. }
    \label{fig:f2_kernel_detail}
\end{figure*}

As seen in Figure~\ref{fig:dominant_terms_p_f2}, the structure of the full dominance plot strongly resembles that of a dominance plot for $\Ftwo$ alone, which itself reflects the behavior of $\Ftwo$ in the $\delta$-$\theta$ plane (Figure~\ref{fig:f2_kernel_detail}). 

Including $P_{ij}$ expands some regions (near their borders, $P_{ij}$ can move the maximum term from just under to just over $5\times$ the next-largest term). In Figure~\ref{fig:dominant_terms_p_f2}, we choose $f=\sqrt{5}$ for the $P_{ij}$ and $\Ftwo$ panels to agree with our choice of $f=5$ for the product $P_{ij}\Ftwo$. As discussed in \S\ref{subsec:regions_of_the_dominance_map} (equation \ref{eqn:meds_product_notequal}), the $P_{ij}$ and $\Ftwo$ dominance criteria cannot simply be multiplied together to determine dominance in $B_{ij}$, but these two contributions independently illuminate the full bispectrum dominance map.

The dynamic range of $\Ftwo$ is larger than that of $P_{ij}$, so the $\Ftwo$ kernel determines most of the dominance map of Figure~\ref{fig:dominance_plot_guide}. The middle term of the $\Ftwo$ kernel, $\left({k_i}/{k_j}+{k_j}/{k_i}\right)(\hat{\bk}_i\cdot\hat{\bk}_j)$ in equation~\pref{eqn:F2}, varies the most between configurations: it can be positive or negative, and can be very large when one side is much smaller than the other (for example, surrounding $\delta=0, \theta/\pi = 1$ in Figure~\ref{fig:f2_kernel_detail}). Alternatively, $\Ftwo$ can approach arbitrarily close to zero (black curves in Figure~\ref{fig:f2_kernel_detail}).

\subsection{Regions of the Dominance Map}
\label{subsec:regions_of_the_dominance_map}
In this section, we step through each region of the dominance plot of Figure~\ref{fig:dominance_plot_guide} from left to right to discuss the dominance behavior. In general, the relative magnitudes of the $B_{12}$, $B_{23}$, and $B_{31}$ differ across configurations due to differences in the $(\delta, \theta)$ dependence of the three wavenumbers $k_1$, $k_2$ and $k_3$ (given in equation~\pref{eqn:k2_k3_fcn_delta_theta} and Figure~\ref{fig:k2_k3_magnitude}).

For each region, we discuss the behavior of the $P_{ij}$ and $\Ftwo_{ij}$ that enter the bispectrum. We build up understanding of each region by first analyzing their behavior separately, then considering the implications for the full dominance plot. We take this approach because the power spectrum products behave very differently from the $\Ftwo$ kernels, even though dominance is determined by the median ratios of terms $B_{ij}^{\rm nw}/B_{jk}^{\rm nw}$, which are medians of products and not products of medians:
\begin{equation}
\label{eqn:meds_product_notequal}
{\rm med} \left\lvert \frac{B_{ij}^{\rm nw}}{B_{jk}^{\rm nw}} \right\rvert = {\rm med} \left\lvert \frac{P_{ij}^{\rm nw}\Ftwo_{ij}}{P_{jk}^{\rm nw}\Ftwo_{jk}} \right \rvert \neq {\rm med} \left\lvert \frac{P_{ij}^{\rm nw}}{P_{jk}^{\rm nw}} \right\rvert \times {\rm med} \left\lvert \frac{\Ftwo_{ij}}{\Ftwo_{jk}} \right \rvert.
\end{equation}

We note that the power spectrum is maximal at $k_{\rm peak}\approx \, 0.015 \hMpc$; above $k_{\rm peak}$, $P^{\rm nw}(k)$ declines monotonically as $1/k$. Since our analysis covers the range $0.01 \leq k_1/[h/{\rm Mpc}] \leq 0.2$, it is a good approximation that in our $k$-range of interest the broadband power spectrum falls as $P(k) \propto 1/k$. This approximation fails only in the low-$\delta$, high-$\theta$ region where $k_3$ can be sufficiently small that $P(k_3)$ increases with $k_3$ (discussed in \S\ref{subsubsec:dominance_green} below).

\subsubsection{Red region, $B_{12}$ dominant} 
\label{subsubsec:dominance_red}
When $\theta/\pi \lesssim 0.5$, Figure~\ref{fig:dominance_plot_guide} shows that $B_{12}$ is the dominant term in the bispectrum cyclic sum \pref{eqn:B0}. In this red region, configurations are constructive where $\PBAO_1$ and $\PBAO_2$ (defined in equation~\ref{eqn:PBAO_wi_def}) are in phase, and destructive where they are out of phase. $B_{12}$ dominates because in both the $\Ftwo$ kernel and the products of power spectra $P_{ij}$, the pre-cyclic terms are largest.

$\Ftwo_{12}$ is much larger than $\Ftwo_{23}$ and $\Ftwo_{31}$ (as shown in Figure~\ref{fig:f2_kernel_detail}) because only in $\Ftwo_{12}$ is the sign of the dipole contribution $(k_i/k_j + k_j/k_i)$ positive. The dot product of the unit vectors $\hat{\bk}_1$ and $\hat{\bk}_2$, which determines the sign of the dipole contribution, approaches $+1$ in $\Ftwo_{12}$. For $\Ftwo_{23}$ and $\Ftwo_{31}$, the relevant dot product instead approaches $-1$. While $\Ftwo_{23}$ and $\Ftwo_{31}$ therefore contain both positive and negative contributions of similar magnitude, all contributions to $\Ftwo_{12}$ are positive, so $\Ftwo_{12}$ will be the largest $\Ftwo_{ij}$. For example, where $\delta$ vanishes as well as $\theta$ (in the lower left corner, $\times$ symbol), two sides, $k_1$ and $k_2$, are equal, while $k_3 = 2k_1$. With these side lengths and dot products, $\Ftwo_{12} = 2$ and $\Ftwo_{31} = \Ftwo_{23} = -0.25$. $\Ftwo_{12}$ exceeds the other two $\Ftwo_{ij}$ by a factor of eight. 

The effect of the power spectrum products is to further separate the three cyclic terms. For $0 \leq \theta/\pi < 0.5$, the triangle is obtuse (see $\theta$ in Figure~\ref{fig:triangle_theta_def}). As $\theta \to 0$ and the triangle fully opens, $k_3$ approaches $k_1 + k_2$. For these obtuse triangles $k_3 > k_2 > k_1$ (see Figure~\ref{fig:k2_k3_magnitude}) because $k_2$ always exceeds $k_1$. The power spectrum is monotonically decreasing, so the $k_i$ ordering implies $P(k_1) > P(k_2) > P(k_3)$. Thus $P_{12} > P_{31} > P_{23}$, reinforcing the order of the $\Ftwo_{ij}$. 

For nonzero $\delta$ (e.g., $+$ symbol in Figure~\ref{fig:dominance_plot_guide}), $B_{31}$ grows with $\delta$, but $B_{12}$ remains dominant by our dominance criterion of $f=5$. For large $\delta$, $k_1$ is small relative to $\delta \lambda_f/2$. The other two wavenumbers $k_2$ and $k_3$ are both larger than $k_1$ (Figure~\ref{fig:k2_k3_magnitude}), so $P(k_3)$ approaches $P(k_2)$. As a result, $P_{31}$ and $P_{12}$ are of similar magnitude (as in the purple region at low $\theta$ and high $\delta$ in the leftmost panel of Figure~\ref{fig:dominant_terms_p_f2}). The magnitudes of the $\Ftwo_{12}$ and $\Ftwo_{31}$ kernels also grow as $\delta$ increases, with the $\Ftwo_{31}$ kernel approaching but remaining smaller than $\Ftwo_{12}$. As $\delta$ continues to increase, $B_{31}$  comes within a factor of 10 of $B_{12}$, causing a purple region to appear in the upper left corner of the right panel of Figure~\ref{fig:dominant_terms}. We note that this effect is too small to appear when the dominance criterion is $f=5$, our choice in the main analysis of this work (as in Figure~\ref{fig:dominance_plot_guide}).

\subsubsection{Middle region, no term dominant (black) or $B_{12}$ and $B_{31}$ dominant (purple)} 
\label{subsubsec:dominance_black_purple}
Around $\theta/\pi=0.6$, at most one term in the cyclic sum can be neglected. In the black region at low $\delta$, all three terms are of comparable magnitude; in the purple region at larger $\delta$, $B_{23}$ shrinks, but $B_{12}$ and $B_{31}$ are still large and of similar magnitude. The black region contains triangles that are nearly equilateral; triangles with $\theta/\pi=2/3$ and $\delta=0$ are equilateral for all $k_1$. Since all three sides and angles are equal, all three terms in the bispectrum are identical, and no term can dominate any other.

In the purple region, $B_{23}$ is negligible compared to $B_{12}$ and $B_{31}$. As $\delta$ increases along the $k_2=k_3$ line shown in Figures~\ref{fig:k2_k3_magnitude} and \ref{fig:dominance_plot_guide}, $k_2$ and $k_3$ grow larger than $k_1$. Since $k_2=k_3$, $B_{12}$ and $B_{31}$ remain equal; their ratio will deviate little from unity. But as $\delta$ increases, both $\Ftwo_{23}$ and $P_{23}$ shrink relative to the other terms. In particular, $\Ftwo_{23}$ approaches zero. For large $\delta$ along $k_2=k_3$, the unit vectors $\hat{\bk}_2$ and $\hat{\bk}_3$ are antiparallel as $k_1$ is relatively small, and their dot product $(\hat{\bk}_2\cdot\hat{\bk}_3)$ becomes $-1$. Since $k_2=k_3$, the dipole contribution $(k_2/k_3 + k_3/k_2) = 2$ in the $\Ftwo_{23}$ kernel, and $\Ftwo_{23}$ therefore vanishes. Furthermore, Figure~\ref{fig:k2_k3_magnitude} shows that at large $\delta$, $k_2$ and $k_3$ are much larger than $k_1$. The product of power spectra $P_{23}$ is therefore smaller than the other two products $P_{12}$ and $P_{31}$, which both involve the much larger $P(k_1)$. Both $\Ftwo_{23}$ and $P_{23}$ shrink as $\delta$ increases, so $B_{23}$ becomes smaller than the other two terms and can be neglected. Thus $B_{12}$ and $B_{31}$ dominate the bispectrum cyclic sum.

\subsubsection{Blue region, $B_{31}$ dominant} 
\label{subsubsec:dominance_blue}
At the right side of Figure~\ref{fig:dominance_plot_guide}, where $\theta$ is large (circle symbol), $B_{31}$ dominates the cyclic sum. Both $\Ftwo_{31}$ and $P_{31}$ are large relative to the other $\Ftwo$ kernels and products of power spectra.

In the $\Ftwo$ kernel, as shown in Figure~\ref{fig:f2_kernel_detail}, $\Ftwo_{12}$ vanishes, and negative contributions to $\Ftwo_{23}$ make it smaller than $\Ftwo_{31}$. $\Ftwo_{12}$ vanishes because in this region, triangles are in the squeezed limit, where $\bk_2 \approx -\bk_1$. The third wavenumber $k_3$ approaches $k_2-k_1$, meaning
\begin{equation}
\label{eqn:k3_thetapi_limit}
    k_3 \to \frac{\delta\lambda_f}{2},
\end{equation}
so $k_3$ is small relative to the other two wavenumbers (see Figure~\ref{fig:k2_k3_magnitude}). At the same time, the dot product $\hat{\bk}_1\cdot\hat{\bk}_2$ approaches $-1$, so $\Ftwo_{12}$ behaves as 
\begin{equation}
    \label{eqn:f2_12_blue}
    \Ftwo_{12} \to 1 - \frac{1}{2}\left(\frac{k_1}{k_2} + \frac{k_2}{k_1}\right) \sim 0.   
\end{equation}
In equation \pref{eqn:f2_12_blue}, the difference $\delta \lambda_f/2$ between $k_1$ and $k_2$ \pref{eqn:delta_theta_def} is much smaller than $k_1$, so $k_1 \sim k_2$ and $\Ftwo_{12}$ vanishes. Meanwhile, the other two terms do not vanish; $\hat{\bk}_2 \cdot \hat{\bk}_3 = -1$ but $\hat{\bk}_1 \cdot \hat{\bk}_3 = +1$, so $\Ftwo_{23}$ and $\Ftwo_{31}$ approach 
\begin{equation}
    \label{eqn:f2_23_blue}
    \Ftwo_{23} \to 1 - \frac{1}{2}\left(\frac{k_2}{k_3} + \frac{k_3}{k_2}\right) \\
\end{equation}
\begin{equation}
    \label{eqn:f2_31_blue}
    \Ftwo_{31} \to 1 + \frac{1}{2}\left(\frac{k_3}{k_1} + \frac{k_1}{k_3}\right).
\end{equation}
The magnitudes of $k_1$ and $k_2$ are comparable, so the dipole contributions $(k_2/k_3 + k_3/k_2)$ in $\Ftwo_{23}$ (equation \ref{eqn:f2_23_blue}) and $(k_3/k_1 + k_1/k_3)$ in $\Ftwo_{31}$ (equation \ref{eqn:f2_31_blue}) are similar in magnitude. Only in $\Ftwo_{31}$ do both contributions have the same sign, so $\Ftwo_{31}$ is larger than $\Ftwo_{23}$.

The $P_{ij}$ reinforce the behavior of the $\Ftwo$ kernel. $k_3$ is small, but still large enough that $P(k_3)$ is monotonically decreasing; for $\delta$ of a few, equation~\pref{eqn:k3_thetapi_limit} is near $k_{\rm Peak}$ (\S\ref{subsec:regions_of_the_dominance_map}). In this limit $k_2 = k_1+k_3$, so $k_2$ must be the largest of the three wavenumbers. In the power spectrum, then, $P(k_2) < P(k_3), P(k_1)$. $P_{12}$ and $P_{23}$, which include $P(k_2)$, are therefore smaller than $P_{31}$, which does not. The largest power spectrum term is therefore $P_{31}$ as in the $\Ftwo$ kernel, and $B_{31}$ dominates the bispectrum.

\subsubsection{Green region, $B_{23}$ and $B_{31}$ dominant} 
\label{subsubsec:dominance_green}
In the lower right corner of Figure~\ref{fig:dominance_plot_guide} around $\delta=0$, $\theta/\pi=1$ (star symbol), the dominant terms are $B_{23}$ and $B_{31}$. Dominance is once again driven by the $\Ftwo$ kernel; $\Ftwo_{23}$ and $\Ftwo_{31}$ are very large (as shown in Figure~\ref{fig:f2_kernel_detail}) while $\Ftwo_{12}$ vanishes. 

$\Ftwo_{12}$ vanishes for the same reason it does in the blue region, but the magnitude of $\Ftwo_{23}$ is more comparable to that of $\Ftwo_{31}$ than it is at higher $\delta$. Along the line where $\theta/\pi=1$, the blue region transitions to green at roughly $\delta \approx 1$ (square symbol in Figure~\ref{fig:dominance_plot_guide}). Here, $k_3 = \lambda_f/2 \approx 0.0285 \hMpc$, which is nearing the lowest $k_1$ in our range ($k_1 = 0.01 \hMpc$). Both $k_2$ and $k_1$ therefore exceed $k_3$ by up to a factor of $10$. As in the blue region, triangles in the green region are squeezed, so the dot products between unit wavevectors are the same as in the blue region. The kernels then behave as equations (\ref{eqn:f2_12_blue}--\ref{eqn:f2_31_blue}), and $\Ftwo_{12}$ (equation \ref{eqn:f2_12_blue}) vanishes. Unlike in the blue region, however, the dipole contribution $(k_i/k_j + k_j/k_i)$ is large enough that the constants in equations \pref{eqn:f2_23_blue} and \pref{eqn:f2_31_blue} become insignificant. $\Ftwo_{23}$ and $\Ftwo_{31}$ are thus both comparably large; $\Ftwo_{31}$ is large and positive ($\hat{\bk}_1\cdot\hat{\bk}_3 = +1$) while $\Ftwo_{23}$ is large and negative ($\hat{\bk}_2\cdot\hat{\bk}_3 = -1$). This region satisfies our double dominance criterion \pref{eqn:double_dominance_criterion}, as even the sum $B_{23}+B_{31}$ exceeds the very small third term $B_{12}$---for example, by a factor of $10^5$ at $\delta=0.01, \theta/\pi=1$.

Our assumption that $P(k)$ falls monotonically with $k$ breaks down in the green region. Because $k_3$ is proportional to $\delta$ (as in equation~\ref{eqn:k3_thetapi_limit}), $k_3$ becomes very small for small $\delta$ (see Figure~\ref{fig:k2_k3_magnitude}). Our previous analysis assumed all wavenumbers were large enough that the power spectrum is monotonically decreasing, but in the green region, $k_3$ can be in the regime where $P(k)$ increases with $k$. Both $k_1$ and $k_2$ are still greater than $k_{\rm Peak}$, in the regime where $P(k)$ falls with $k$. For very small $k_3$, then, $P(k_3)$ may be smaller than $P(k_1)$ and $P(k_2)$. As a result, $P_{12}$ can be the largest product of power spectra, despite the fact that $k_1$ and $k_2$ are much larger than $k_3$. However, the power spectrum is overshadowed by the behavior of the $\Ftwo$ kernel. Even for $\delta=0.01$, where $P_{12}$ can exceed the other two products of power spectra by a factor of 25 at small $k_1$, the median ratio of power spectra products, ${\rm med}\left[P_{12}/P_{23}\right]$, is only 5. But the median ratio of $\Ftwo$ kernels, ${\rm med}\left[\Ftwo_{12}/\Ftwo_{23}\right]$, is of order $10^{-8}$ because $k_1$ so nearly equals $k_2$, driving $\Ftwo_{12}$ to vanish. The dominance structure is thus driven primarily by the $\Ftwo$ kernel.

\subsection{Ordering of Subdominant Terms}
\label{subsec:ordering_of_subdominant_terms}

\begin{figure}
\includegraphics{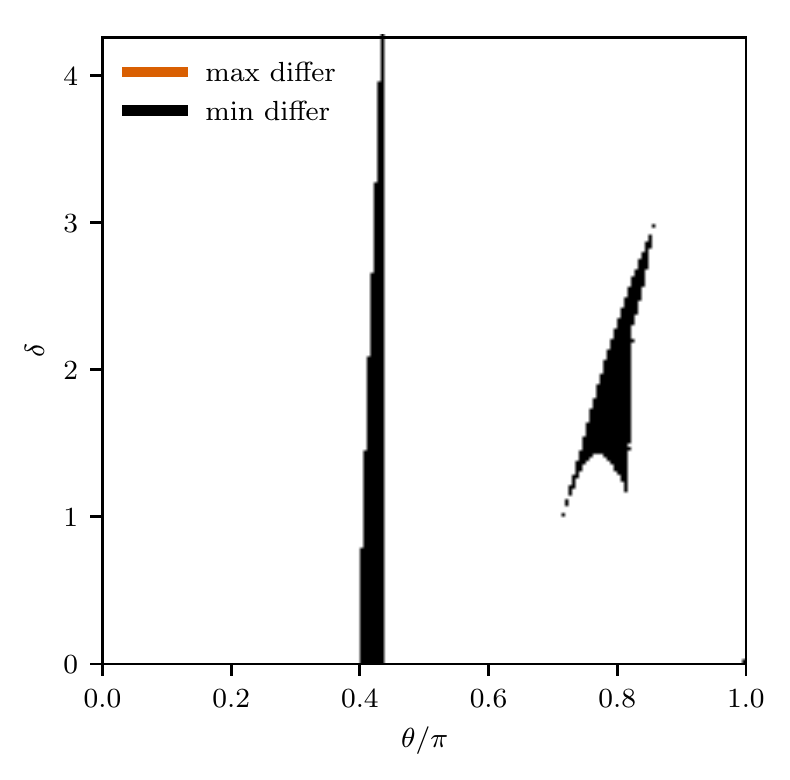}
\caption{\label{fig:max_min_term_differs} The dominant term in the cyclic sum composing the bispectrum \pref{eqn:B0} is determined by both $P_{ij}$ and $\Ftwo$ (as shown in Figure~\ref{fig:dominant_terms_p_f2}), and the two almost always act in the same direction. The maximum term never differs between $P_{ij}$ and $\Ftwo$ (orange regions, none shown), but the minimum terms are swapped in two regions (black). In these regions, discussed in \S\ref{subsec:ordering_of_subdominant_terms}, the term with minimum $P_{ij}$ has the second-largest $\Ftwo_{ij}$.}
\end{figure}

\begin{figure}
\includegraphics{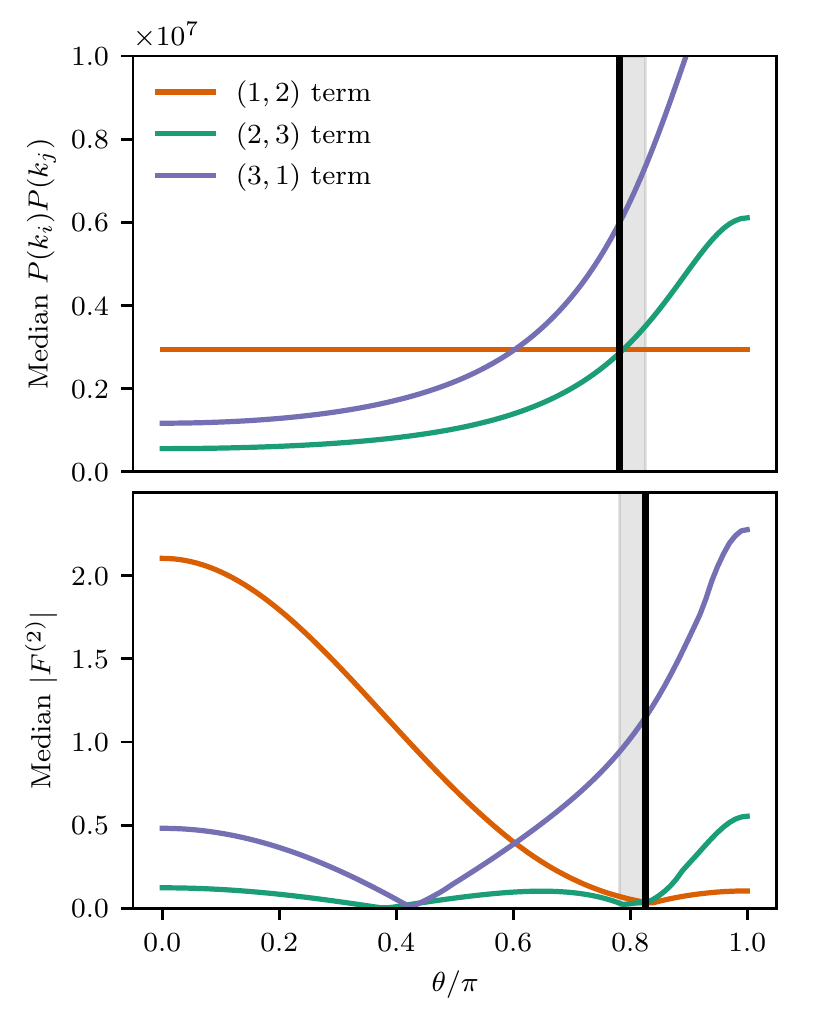}
\caption{\label{fig:medians_F2_p} In the shaded region, the ordering of subdominant terms differs between $P_{ij}$ and $\Ftwo_{ij}$: $P_{12}$ is the smallest $P_{ij}$, while the minimum $\Ftwo_{ij}$ is $\Ftwo_{23}$ (compare Figure~\ref{fig:max_min_term_differs}). As discussed in \S\ref{subsec:ordering_of_subdominant_terms}, this region arises due to differences in the behavior of the median between the power spectrum and the $\Ftwo$ kernel. Around $\theta/\pi \sim 0.6$, both $\Ftwo_{12}$ and $\Ftwo_{31}$ are positive for all $k_1$, so their medians cross at the same $\theta$ as the medians of $P_{12}$ and $P_{31}$. At $\theta/\pi = 0.78$, however, $P_{23}$ crosses above $P_{12}$ (solid vertical line in top panel); $\Ftwo_{23}$ lags behind, crossing above $\Ftwo_{12}$ at $\theta/\pi=0.83$ (solid vertical line in bottom panel). The ordering also differs around $\theta/\pi \sim 0.4$, as further discussed in \S\ref{subsec:ordering_of_subdominant_terms}. }
\end{figure}

The $B_{ij}$ of equation \pref{eqn:B0} with maximum median $P_{ij}$ is everywhere also the term with maximum median $|\Ftwo_{ij}|$. The ordering of terms differs only in the regions shown in Figure~\ref{fig:max_min_term_differs}, where the two subdominant terms are swapped. These regions arise because the $\Ftwo$ kernel can be either positive or negative. In the range of wavenumbers of interest, the power spectrum is always positive but monotonically decreasing, so the products $P_{ij}$ change smoothly. $P_{23}$ and $P_{31}$, smaller than $P_{12}$ for small $\theta$ (see the left panel of Figure~\ref{fig:dominant_terms_p_f2} and the top panel of Figure~\ref{fig:medians_F2_p}), cross above $P_{12}$ around the equilateral configuration ($\theta/\pi=0.67$). $P_{31}$ is the first to cross above $P_{12}$ because $k_2 \geq k_1$, so $P_{31} \geq P_{23}$. $P_{23}$ lags behind (see the top panel of Figure~\ref{fig:medians_F2_p}). 

The behavior of the $\Ftwo$ kernel (equation \ref{eqn:F2}) is not as simple, as shown in the lower panel of Figure~\ref{fig:medians_F2_p}. First, it can be either positive or negative (see Figure~\ref{fig:f2_kernel_detail}), explaining the $\theta/\pi \sim 0.4$ region in Figure~\ref{fig:max_min_term_differs} where the ordering of the subdominant terms differs between $\Ftwo$ and $P_{ij}$. The difference arises because the absolute values of the two subdominant terms $\Ftwo_{23}$ and $\Ftwo_{31}$ spuriously cross when both are small. In detail, we take the absolute value of each $\Ftwo_{ij}$, since the magnitudes of each term, not their signs, determine dominance. For $\theta \sim 0$, both $\Ftwo_{23}$ and $\Ftwo_{31}$ are negative, with $\Ftwo_{31}$ more negative than $\Ftwo_{23}$. For $\theta/\pi \gtrsim 0.4$, both terms are positive, with $\Ftwo_{31}$ more positive than $\Ftwo_{23}$. $\Ftwo_{31}$ must therefore cross above $\Ftwo_{23}$, and it does so in the same region around $\theta/\pi \sim 0.4$ where both terms cross zero---but the terms are not equal to zero where they cross each other. As shown in Figure~\ref{fig:f2_kernel_detail}, the value of $\theta$ at which $\Ftwo_{ij}$ crosses zero depends on the ratio between the two sides $k_i$ and $k_j$, so $\Ftwo_{23}$ becomes positive at slightly lower $\theta$ than does $\Ftwo_{31}$. As $\Ftwo_{31}$ approaches zero, its absolute value falls below the small and positive $\Ftwo_{23}$ at $\theta/\pi=0.42$ (with $\delta=2.1$, for example, as in Figure~\ref{fig:medians_F2_p}). After $\Ftwo_{31}$ becomes positive, it crosses $\Ftwo_{23}$ at $\theta/\pi = 0.44$. Meanwhile in the product of power spectra, the median $P_{31}$ exceeds the median $P_{23}$ for all $\theta$. Therefore, in this narrow region where the median $\Ftwo_{31}$ falls below the median $\Ftwo_{23}$, the smallest $P_{ij}$ term is not the smallest $\Ftwo_{ij}$ term.

The order of the subdominant terms also differs around $\theta/\pi \sim 0.8$ (Figure~\ref{fig:max_min_term_differs}). Figure~\ref{fig:medians_F2_p} shows that this region arises due to differences in the behavior of the median between the power spectrum and the $\Ftwo$ kernel. The power spectrum decreases monotonically, so the median $P_{ij}$ occurs at the median $k_1$. Therefore $P_{23}$ and $P_{12}$ are equal when $k_3=k_1$ is evaluated at the median $k_1$ (see equation~\ref{eqn:k1_k3_equal_delta}). In Figure~\ref{fig:medians_F2_p}, $P_{23}$ crosses above $P_{12}$ at $\theta/\pi=0.78$ (with $\delta=2.1$). Unlike the power spectrum, the $\Ftwo$ kernel can be positive or negative, and in some configurations it is positive for some values of $k_1$ and negative for others. For these configurations, the absolute value of $\Ftwo$ is not a monotonic function of $k_1$, so its median does not necessarily occur at the median $k_1$. Therefore $\Ftwo_{23}$ crosses above $\Ftwo{12}$ at $\theta$ such that the $k_3$ that corresponds to the median $\Ftwo_{23}$ equals the $k_1$ that corresponds to the median $\Ftwo_{12}$. In the example of Figure~\ref{fig:medians_F2_p}, the solution is $\theta/\pi=0.83$. $P_{12}$ becomes the minimum $P_{ij}$ at $\theta/\pi=0.78$ while $\Ftwo_{12}$ does not become the minimum $\Ftwo_{ij}$ until $\theta/\pi=0.83$. Therefore in the shaded region of Figure~\ref{fig:medians_F2_p} between these two crossings ($0.78 \leq \theta/\pi \leq 0.83$), the order of the subdominant terms differs.

In contrast, around $\theta/\pi \sim 0.6$, both $\Ftwo_{12}$ and $\Ftwo_{31}$ are positive for all $k_1$. Their medians are both found at the median value of $k_1$, so $\Ftwo_{31}$ crosses above $\Ftwo_{12}$ where $k_2=k_3$ (equation \ref{eqn:k2_k3_equal_delta}, evaluated at the median $k_1$). As the median $P_{ij}$ occurs at the median $k_1$, $P_{31}$ also crosses above $P_{12}$ where equation \pref{eqn:k2_k3_equal_delta} is evaluated at the median $k_1$. $\Ftwo_{31}$ therefore becomes the maximum $\Ftwo_{ij}$ at the same value of $\theta$ where $P_{31}$ becomes the maximum $P_{ij}$.

Though we set out to explain the regions where the minimum $\Ftwo_{ij}$ differs from the minimum $P_{ij}$, our analysis also explains why the largest $P_{ij}$ is always also the largest $\Ftwo_{ij}$ (see Figure~\ref{fig:max_min_term_differs}). The median behaves most simply for $\Ftwo_{12}$ and $P_{12}$, which are both the maximum term at low $\theta$. As $\theta$ increases, $P_{31}$ is always the first to cross $P_{12}$, and $\Ftwo_{31}$ is always the first to cross $\Ftwo_{12}$. $\Ftwo_{23}$ and $P_{23}$ are never the maximum term because $k_2 \geq k_1$. The complexity of the ordering of subdominant terms arises from the $\Ftwo_{23}$ and $P_{23}$ terms, but because these terms are never the maximum terms, the maximum term never differs between $P_{ij}$ and $\Ftwo_{ij}$. 

\section{Decomposition into Eigen-Root-Mean-Square Plots}
\label{sec:analytic_eigen}
We now show how the RMS amplitude $\mathcal{A}$ (equation \ref{eqn:mathcalA}) can frequently be approximated as a linear combination of three terms. We first require an expression for the ratio $R$ (equation \ref{eqn:R_bispec}) of the full bispectrum \pref{eqn:B0} to its no-wiggle analog. We wish to leverage the fact that the BAO are a small fractional feature in the power spectrum, so we write
\begin{align}
P_{ij} = P^{\rm nw}_{ij}[1 + w_{ij}]
\end{align}
with $P^{\rm nw}_{ij}$ the product of two no-wiggle power spectra and $w_{ij}$ the BAO feature in the product $P_{ij}$ \pref{eqn:pij_def} of linear power spectra. In particular we split the matter transfer function $T_{\rm m}$ into smooth and oscillatory pieces as 
\begin{equation}
T_{\rm m}(k)= T_{\rm sm}(k)+\omega(k)j_0(k\tilde{s}), 
\end{equation}
where $T_{\rm sm}(k)$ and $\omega(k)$ are smooth functions of $k$ \citep{EisensteinHu:98}, $\omega$ is small (because $\Omega_{\rm b}/\Omega_{\rm m} \ll 1$), and $j_0(x) = \sin(x)/x$ is the order zero spherical Bessel function. 

The power spectrum is proportional to the primordial power spectrum $P_{\rm pri}(k)$ and the matter transfer function as
\begin{equation}
P(k) = P_{\rm pri}(k) T_{\rm m}^2(k).
\end{equation}
We suppress the redshift dependence of the power spectrum for simplicity, as it does not affect our analysis. The products of power spectra are then
\begin{align}
    P_{ij} = & P_{\rm pri}(k_i)P_{\rm pri}(k_j)T_{\rm sm}^2(k_i)T_{\rm sm}^2(k_j)\; \nonumber \\
    & \times \left[ 1+ \frac{\omega(k_i) j_0(k_is)}{T_{\rm sm}(k_i)} \right]^2 \left[ 1+ \frac{\omega(k_j) j_0(k_js)}{T_{\rm sm}(k_j)} \right]^2.
\end{align}
Taylor-expanding the fractions $\omega(k_i) j_0(k_is)/T_{\rm sm}(k_i)$ to leading order in $\omega$ we have
\begin{align}
\label{eqn:wij_omega_tsm}
w_{ij} = \frac{P_{ij}}{P^{\rm nw}_{ij}} -1 \approx 2\left[\frac{\omega(k_i) j_0(k_is)}{T_{\rm sm}(k_i)} + \frac{\omega(k_j) j_0(k_js)}{T_{\rm sm}(k_j)}\right],
\end{align}
where we used the fact that the no-wiggle power spectrum is simply $P^{\rm nw}(k) = P_{\rm pri}(k)T_{\rm sm}^2(k)$.

In the remainder of this section, we show that in regions where only one or two terms dominate the bispectrum, the variance $\mathcal{A}^2$ of the full bispectrum is approximated to leading order by the variance of only the dominant term or terms. We first consider the case where one term is dominant and the other two negligible, and we then consider the case where one term is negligible and the other two must be retained.

\subsection{Single Term Dominant}
\label{subsec:analytic_eigen_singleterm}
We first consider the case where one term in the bispectrum dominates over the other two; without loss of generality we take this to be the first term. 

We calculate the RMS amplitude $\mathcal{A}$ (equation \ref{eqn:mathcalA}) from the variance of the ratio $R$, defined in equation \pref{eqn:R_bispec} as
\begin{align}
R = \frac{B_{12} + B_{23} + B_{31}}{B_{12}^{\rm nw} + B_{23}^{\rm nw} + B_{31}^{\rm nw}}.
\label{eqn:ratio_fullbispec}
\end{align}
Our goal is to show that the variance $\mathcal{A}_{12}^2$ of the approximate ratio 
\begin{align}
R_{12} = \frac{B_{12}}{B_{12}^{\rm nw}} = 1 + w_{12} 
\label{eqn:R12}
\end{align}
is the same as that of the full ratio, $\mathcal{A}^2$, to leading order in one or the other of two small parameters we define, 
\begin{align}
\epsilon_{ij} \equiv \frac{B_{ij}}{B_{12}},\;\;\;
\epsilon_{ij}^{\rm nw} \equiv\frac{B_{ij}^{\rm nw}}{B_{12}^{\rm nw}}.
\label{eqn:eps_defns}
\end{align}
We first notice that the variance of $R_{12}$ is 
\begin{align}
\mathcal{A}_{12}^2 = \left<w_{12}^2 \right> - \left<w_{12} \right>^2;   
\label{eqn:lo_variance}
\end{align}
the constant term in equation \pref{eqn:R12} of course contributes no variance. Since $\mathcal{A}_{12}^2$ is second-order in the small parameter $w_{12}$, we neglect all corrections at third order and higher. We will find that the difference between the full variance and the variance of $R_{12}$ vanishes at second order.

Factoring out the dominant term in the numerator and denominator of equation \pref{eqn:R12} and using the definitions above, we have the ratio as
\begin{align}
R = & \left(\frac{B_{12}}{B_{12}^{\rm nw}}\right)\frac{1 + \epsilon_{23} + \epsilon_{31}}{1 + \epsilon_{23}^{\rm nw} + \epsilon_{31}^{\rm nw}}\nonumber\\
\approx & R_{12}\left(1 + \epsilon_{23} + \epsilon_{31} \right) \left[1 - \epsilon_{23}^{\rm nw} - \epsilon_{31}^{\rm nw}  + (\epsilon_{23}^{\rm nw} + \epsilon_{31}^{\rm nw})^2 \right]\nonumber\\
 = &R_{12}\left[1 + (\epsilon_{23}- \epsilon_{23}^{\rm nw}) + (\epsilon_{31} - \epsilon_{31}^{\rm nw})\right.\nonumber\\  
&\left. - (\epsilon_{23} + \epsilon_{31})(\epsilon_{23}^{\rm nw} + \epsilon_{31}^{\rm nw}) + (\epsilon_{23}^{\rm nw} + \epsilon_{31}^{\rm nw})^2 + \mathcal{O}(\epsilon^3) \right]. 
\label{eqn:approx_ratio}
\end{align}
In the second, approximate equality, we have Taylor-expanded the denominator to second order in $\epsilon$. We include the second-order term for the moment but see that it drops out of our end result.

We now seek to exploit the fact that the BAO feature itself is small, i.e. $w_{ij} \ll 1$ (equation \ref{eqn:wij_omega_tsm} with small $\omega(k)$). Using $w_{ij}$ as defined in equation \pref{eqn:Rij_wij_def} to simplify differences of $\epsilon_{ij}$ and $\epsilon_{ij}^{\rm nw}$, we have
\begin{align}
\epsilon_{23} - \epsilon_{23}^{\rm nw} &= \frac{B_{23}^{\rm nw}}{B_{12}^{\rm nw}}\left[\frac{1 + w_{23}}{1 + w_{12}} -1 \right]\nonumber\\
&\approx \epsilon_{23}^{\rm nw}(w_{23} - w_{12} + w_{12}^2 - w_{12}w_{23})\nonumber\\
&=  \epsilon_{23}^{\rm nw}\Delta^w_{23, 12}(1 - w_{12}),
\label{eqn:eps_diff}
\end{align}
where to obtain the first equality we substituted the definitions \pref{eqn:eps_defns} and to obtain the second we Taylor-expanded to leading order in $w$. In the third equality, we defined 
\begin{align}
\Delta^w_{23, 12} = w_{23} - w_{12},    
\end{align}
which is $\mathcal{O}(w)$.
The analog of equation \pref{eqn:eps_diff} holds for the $31$ term by switching $23$ to $31$ everywhere.

We know the variance at leading order is $\mathcal{O}(w^2)$ from equation~\pref{eqn:lo_variance}, so we only retain terms that are second order in a combination of $\epsilon$ and $w$. Our approximate expression for the ratio $R$ is now
\begin{align}
R \approx R_{12} &\left[1 + \epsilon_{23}^{\rm nw} \Delta^w_{23,12} +  \epsilon_{31}^{\rm nw}\Delta^w_{31,12}- (\epsilon_{23} + \epsilon_{31})(\epsilon_{23}^{\rm nw} + \epsilon_{31}^{\rm nw})\right.\nonumber\\
 & +\left.(\epsilon_{23}^{\rm nw} + \epsilon_{31}^{\rm nw})^2 \right].
\label{eqn:full_R_approx_long}
\end{align}
The first term is $\mathcal{O}(1)$, the second and third $\mathcal{O}(\epsilon w)$, and the fourth and fifth $\mathcal{O}(\epsilon^2)$. These last two terms cancel each other to second order; $\epsilon_{ij}-\epsilon_{ij}^{\rm nw}$ (equation \ref{eqn:eps_diff}) is itself second order, so at leading order the second to last term is equal to the last. Equation \pref{eqn:full_R_approx} then simplifies:
\begin{align}
R \approx R_{12} \left[1 + \epsilon_{23}^{\rm nw} \Delta^w_{23,12} +  \epsilon_{31}^{\rm nw}\Delta^w_{31,12} \right].
\label{eqn:full_R_approx}
\end{align}
Now computing the expectation value of $R$ and factoring out $\left<R_{12} \right>$ to enable further Taylor expansions, we find
\begin{align}
\left< R \right> \approx  \left< R_{12} \right> & \left\{1 +  \left< R_{12} \right>^{-1}\left<R_{12}\epsilon_{23}^{\rm nw} \Delta^w_{23,12}\right> \right.\nonumber \\
&\left.+ \left< R_{12} \right>^{-1} \left<R_{12}\epsilon_{31}^{\rm nw}\Delta^w_{31,12} \right> \right\}.
\end{align}
We now square the form above and multiply out to find
\begin{align}
\left< R\right>^2 \approx    \left< R_{12}\right>^2 + 2 & \left< R_{12}\right> \nonumber \\
&\times \left\{ \left<R_{12}\epsilon_{23}^{\rm nw}\Delta^w_{23,12}\right>
+  \left<R_{12}\epsilon_{31}^{\rm nw}\Delta^w_{31,12}\right> \right\}.
\label{eqn:ratio_approx_exp_sq}
\end{align}
Now using equation \pref{eqn:full_R_approx} to compute the expectation value of $R^2$, we obtain 
\begin{align}
\left<R^2 \right> \approx & \left<R_{12}^2 \right> + 2\left\{ \left< R_{12}^2 \epsilon_{23}^{\rm nw}\Delta^w_{23,12} \right>
 +  \left< R_{12}^2 \epsilon_{31}^{\rm nw}\Delta^w_{31,12} \right> \right\}.
\label{eqn:ratio_approx_sq_exp}
\end{align}
The variance of $R$ is then
\begin{align}
\left<R^2\right> -\left<R\right>^2   \approx & \left<R_{12}^2 \right> - \left< R_{12}\right>^2 \\
+ 2 &\bigg\{\left< R_{12}^2 \epsilon_{23}^{\rm nw} \Delta^w_{23,12} \right> - \left<R_{12}\right> \left< R_{12} \epsilon_{23}^{\rm nw}\Delta^w_{23,12}\right> \nonumber \\
&+\left< R_{12}^2 \epsilon_{31}^{\rm nw} \Delta^w_{31,12}\right> -\left<R_{12}\right> \left< R_{12} \epsilon_{31}^{\rm nw} \Delta^w_{31,12}\right>\bigg\}. \nonumber
\end{align}
Recalling that $R_{12} = 1+w_{12}$ (equation~\ref{eqn:R12}) and denoting the variance of $R_{12}$ as $\mathcal{A}_{12}^2$, we find
\begin{align}
\label{eqn:sigma_R_12_difference}
\mathcal{A}^2 & - \mathcal{A}^2_{12} \approx \\
2\, \times &\left\{\left<\left(1+w_{12}\right)^2\epsilon^{\rm nw}_{23}\Delta^w_{23, 12}\right> - \left<1+w_{12}\right>\left< \left(1+w_{12}\right)\epsilon^{\rm nw}_{23}\Delta^w_{23, 12}\right>\right.\nonumber\\
+ & \left.\left<\left(1+w_{12}\right)^2\epsilon^{\rm nw}_{31}\Delta^w_{31, 12}\right> - \left<1+w_{12}\right>\left< \left(1+w_{12}\right)\epsilon^{\rm nw}_{31}\Delta^w_{31, 12}\right>\right\}.\nonumber 
\end{align}
To second order, the difference \pref{eqn:sigma_R_12_difference} cancels. Thus, the difference between $\mathcal{A}^2$ and $\mathcal{A}_{12}^2$ is suppressed by one order relative to $\mathcal{A}_{12}^2$. Therefore $\mathcal{A}_{12}^2$ is the leading contribution to the variance $\mathcal{A}^2$. When a single term dominates the bispectrum cyclic sum, the variance of that single term is a good approximation of the variance of the full bispectrum. In \S\ref{sec:eigenvariances}, we use this fact to better understand the behavior of the RMS map (Figure~\ref{fig:full_colorplot}) in regions dominated by a single term.

\subsection{Double Dominance}
Our goal is to show that if the sum of two terms dominates the third in $R$, then the variance of $R$, $\mathcal{A}^2$, is well approximated by that of the two dominant terms, and that the error in making this approximation is one order higher than the result itself. The requirement that the sum of two terms is much larger than the remaining term is only one condition of our double dominance criterion \pref{eqn:double_dominance_criterion}, but this condition is sufficient to show that $\mathcal{A}$ is well approximated by the contribution of only two terms. Our double dominance criterion is more strict in order to distinguish regions where two terms are both large from those where one term nearly dominates the full bispectrum, and its sum with either of the other two (comparably small) terms is much larger than the remaining term.

Without loss of generality we take $B_{31} \ll B_{12} + B_{23}$. We again begin from equation~\pref{eqn:ratio_fullbispec} and calculate $\mathcal{A}^2$ in terms of the variance of the two dominant terms, $\mathcal{A}^2_{12+23}$. 
We have
\begin{align}
\label{eqn:R_DD}
&R = \left( \frac{ B_{12} + B_{23} } {B_{12}^{\rm nw} + B_{23}^{\rm nw}} \right) \frac{1 + B_{31}/(B_{12} + B_{23})} {1 + B^{\rm nw}_{31}/ ( B^{\rm nw}_{12} + B^{\rm nw}_{23})}\\
& \approx R_{12 + 23} \left(1 + \epsilon_{31/(12 + 23)} \right) \left[ 1 - \epsilon^{\rm nw}_{31/(12 + 23)}  + \left( \epsilon^{\rm nw}_{31/(12 + 23)} \right)^2\right]\nonumber
\end{align}
where to obtain the second line we Taylor-expanded the denominator to second order in $\epsilon \ll 1$. We defined $\epsilon$ and its no-wiggle analog as
\begin{align}
\epsilon_{31/(12 + 23)} = \frac{B_{31}}{B_{12} + B_{23}},\;\;\;\epsilon^{\rm nw}_{31/(12 + 23)} = \frac{B^{\rm nw}_{31}}{B^{\rm nw}_{12} + B^{\rm nw}_{23}}.
\end{align}
We also defined $R_{12 + 23}$ as the first factor in the first line of equation (\ref{eqn:R_DD}), as in equation \pref{eqn:R_ijjkdef}. Multiplying out equation (\ref{eqn:R_DD}) and dropping $\mathcal{O}(\epsilon^3)$ terms, we obtain
\begin{align}
\label{eqn:DD_R_deltaeps}
&R\approx R_{12+23}\left[1 + \Delta^{\epsilon} - \Delta^{\epsilon \times} \right],\nonumber\\
&\Delta^{\epsilon} \equiv \epsilon_{31/(12+23)} - \epsilon^{\rm nw}_{31/(12+23)},\nonumber\\
&\Delta^{\epsilon \times} \equiv \epsilon^{\rm nw}_{31/(12+23)}\epsilon_{31/(12+23)} - \left(\epsilon_{31/(12+23)}^{\rm nw} \right)^2.
\end{align}
The two small parameters $\epsilon_{31/(12+23)}$ and $\epsilon_{31/(12+23)}^{\rm nw}$ differ only in the BAO feature, which we again parameterize by $w_{ij}$ (equation \ref{eqn:Rij_wij_def}):
\begin{align}
\label{eqn:eps_bar_w}
&\epsilon_{31/(12+23)} = B_{31}^{\rm nw}\frac{1 + w_{31}}{B_{12}^{\rm nw} + B_{23}^{\rm nw}}
\left[1 + \frac{B_{12}^{\rm nw}w_{12} + B_{23}^{\rm nw} w_{23}}{B_{12}^{\rm nw} + B_{23}^{\rm nw}}\right]^{-1}  
\end{align}
motivating us to define
\begin{align}
\bar{w}_{12, 23} = \frac{B_{12}^{\rm nw}w_{12} + B_{23}^{\rm nw}w_{23}}{B_{12}^{\rm nw} + B_{23}^{\rm nw}} \equiv \bar{w},    
\end{align}
where in the second, identical equality we are simply noting that we will drop the subscripts on $\bar{w}$. Physically, $\bar{w}$ is the weighted average of the BAO features in the $B_{12}$ and $B_{23}$ terms in the bispectrum \pref{eqn:B0}. Expanding equation (\ref{eqn:eps_bar_w}) to second order in $w$ and $\bar{w}$, we find
\begin{align}
\epsilon_{31/(12 + 23)} \approx \epsilon_{31/(12 + 23)}^{\rm nw} \left[1+ (w_{31} - \bar{w}) + \bar{w}^2 - w_{31} \bar{w} \right].   
\end{align}
So we see that
\begin{align}
&\Delta^{\epsilon} \approx \epsilon_{31/(12+23)}^{\rm nw}(w_{31} - \bar{w})(1-\bar{w}),\nonumber\\
&\Delta^{\epsilon \times} \approx \left(\epsilon_{31/(12+23)}^{\rm nw} \right)^2 (w_{31} - \bar{w})(1-\bar{w}).
\end{align}
Retaining only terms at second order and lower, we then have
\begin{align}
\label{eqn:DD_delta_eps_times}
&\Delta^{\epsilon} \approx \epsilon_{31/(12+23)}^{\rm nw}\left[ w_{31} - \bar{w}\right],\nonumber\\
&\Delta^{\epsilon \times} \sim \mathcal{O}(\epsilon^2 w)
\end{align}
Now we have that to second order $R=R_{12+23}\left(1 + \Delta^\epsilon\right)$, and we find that
\begin{align}
 &\left<R^2 \right> \approx    \left<R^2_{12 + 23} \right> + 2\left< R_{12+23}^2 \Delta^\epsilon \right>,\nonumber\\ 
 &\left< R\right>^2 \approx \left< R_{12+23}\right>^2 + 2\left< R_{12+23}\right>\left<R_{12+23}
\Delta^\epsilon \right>
\end{align}
including all terms at second order and below.

Thus to second order the variance $\mathcal{A}^2_{12+23}$ in the dominant terms differs from the variance $\mathcal{A}^2$ in the full bispectrum only by 
\begin{align}
\label{eqn:DD_sigmaR_sigma1223_diff}
\mathcal{A}^2 - \mathcal{A}^2_{12+23} \approx
2 \left\{\left< R_{12+23}^2 \Delta^\epsilon \right> - \left< R_{12+23}\right>\left<R_{12+23}\Delta^\epsilon \right>\right\}. 
\end{align}
As in \S\ref{subsec:analytic_eigen_singleterm} above, the leading contribution to the difference has $R_{12+23} \approx 1$, in which case the two terms on the right side of equation~\pref{eqn:DD_sigmaR_sigma1223_diff} cancel. The error is therefore $\mathcal{O}\left(\left(\epsilon w\right)^{3/2}\right)$, where our notation $(\epsilon w)^{3/2}$ indicates that the error is third order in small quantities but can have any combination of $\epsilon$ and $w$ reaching that order. In contrast, the result $\mathcal{A}^2_{12+23}$ is second-order in $w$. Thus we have shown that the error of approximating the variance of the full ratio $R$ by that of the ratio of the first two terms, $R_{12+23}$, is one order smaller than the variance itself.

\section{Numerical Eigen-Root-Mean-Square Calculations}
\label{sec:eigenvariances}
\begin{figure*}
\includegraphics{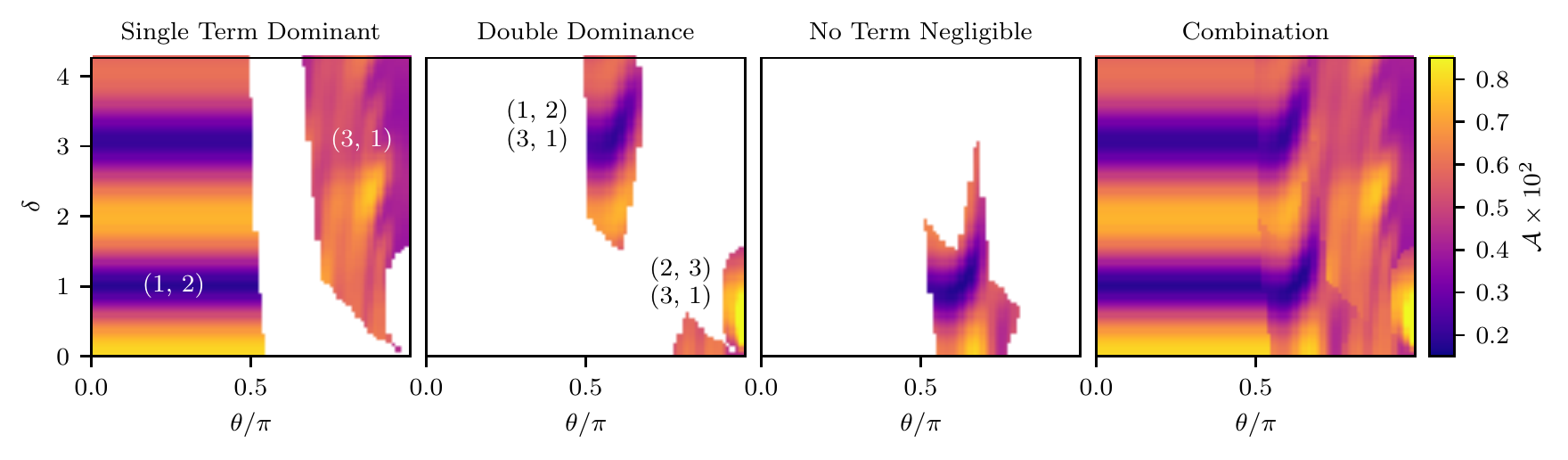}
\caption{\label{fig:eigenplots_one_negligible} The detailed structure of the full RMS map can be understood by considering the RMS amplitude produced by only single terms or pairs of terms in the bispectrum cyclic sum. \emph{Left}---The single-term-dominant contribution: BAO amplitude associated with $R_{12}$ and $R_{31}$ in regions where only $B_{12}$ or $B_{31}$ (indicated on the plot) dominates, detailed in \S\ref{subsec:single_term_variance}. \emph{Middle Left}---The double-term-dominant contribution: regions where one term is negligible, detailed in \S\ref{subsec:double_dominance_variance} (upper middle is $B_{12}+B_{31}$ dominant; lower right is $B_{23}+B_{31}$ dominant). \emph{Middle Right}---The no-term-negligible contribution: regions where all terms are of comparable magnitude, detailed in \S\ref{subsec:eigenvariance_no_term_negligible}. \emph{Right}---By combining the other three panels, we reproduce the full RMS map of Figure~\ref{fig:full_colorplot}. }
\end{figure*}

As shown analytically in \S\ref{sec:analytic_eigen}, in the regions of single and double dominance identified in \S\ref{sec:dominance_regions}, the RMS amplitude of BAO in the full bispectrum simplifies to the RMS amplitude of BAO in the dominant term or terms of the bispectrum cyclic sum \pref{eqn:B0}. We calculate RMS maps for each single term $B_{ij}$ and pair of terms $B_{ij}+B_{jk}$. In Figure~\ref{fig:eigenplots_one_negligible}, we combine the single- and double-term RMS maps in the corresponding single- and double-dominance regions; the result matches the full RMS map of Figure~\ref{fig:full_colorplot} reasonably well. The simplified maps of $\mathcal{A}$ therefore provide a good approximation to the full RMS map. In the remainder of this section, we fully detail the RMS maps for each single term (\S\ref{subsec:single_term_variance}) and pair of terms (\S\ref{subsec:double_dominance_variance}) over the full $(\delta, \theta)$ plane.

\subsection{Single Term Dominant}
\label{subsec:single_term_variance}
While the BAO amplitude in the full bispectrum \pref{eqn:B0} is a complicated function of triangle configuration, many configurations have only a single term dominant, as discussed in \S\ref{sec:dominance_regions}. In those regions, the behavior of the RMS amplitude can be understood through the interaction between pairs of oscillating power spectra. In the red ($\theta \lesssim 0.5\pi$) and blue ($\theta/\pi \sim 1$) regions of the dominance map (Figure~\ref{fig:dominance_plot_guide}) respectively, $B_{12}$ and $B_{31}$ dominate. We expect that in these regions, the RMS map (Figure~\ref{fig:full_colorplot}) is well approximated by the RMS amplitude of BAO in the dominant term only (left panel of Figure~\ref{fig:eigenplots_one_negligible}).

\begin{figure*}
\includegraphics{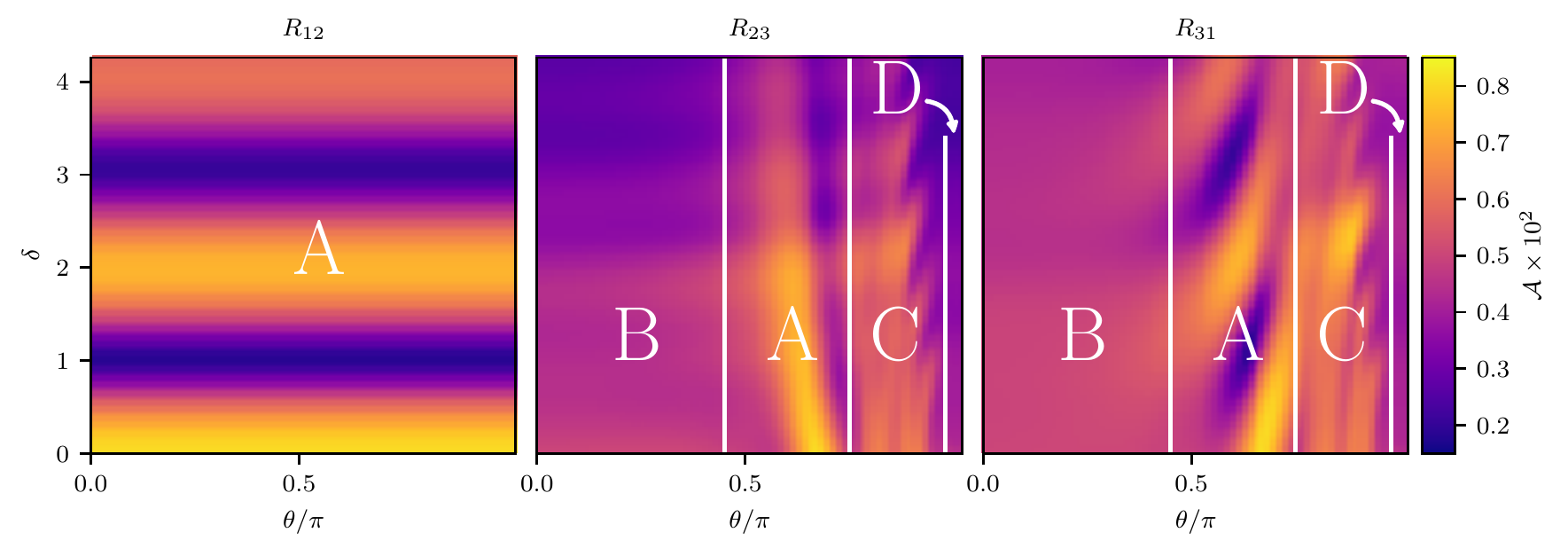}
\caption{\label{fig:k31_term_regions_map} In each labeled region of the single-term-dominant RMS maps, the RMS amplitude $\mathcal{A}$ of the BAO feature is driven by a different mechanism. $\mathcal{A}$ is shown for $R_{12}$ (left panel), $R_{23}$ (middle panel), and $R_{31}$ (right panel). The mechanisms are discussed in detail in \S\ref{subsec:single_term_variance}: interference in region A (\S\ref{subsubsec:eigenvariance_interference}), incoherence in region B (\S\ref{subsubsec:eigenvariance_incoherent}), feathering in region C (\S\ref{subsubsec:eigenvariances_feathering}), and single power spectrum in region D (\S\ref{subsubsec:eigenvariances_singlepk}). The labeled regions are identical for $R_{23}$ and $R_{31}$, while interference is the only mechanism in $R_{12}$. }
\end{figure*}

Figure~\ref{fig:k31_term_regions_map} zooms in on the RMS maps for each of the single terms, that is, the RMS amplitude \pref{eqn:mathcalA} of the ratio of each term to its no-wiggle analog \pref{eqn:Rij_def}. No region with $\delta>0$ has $B_{23}$ dominant (see \S\ref{sec:dominance_regions}), but we discuss this term as well, both for completeness and because it nonetheless shares interesting physics with $B_{31}$.

Each labeled region of the single-term-dominant RMS maps (Figure~\ref{fig:k31_term_regions_map}) is driven by one of the following mechanisms: interference (region B, \S\ref{subsubsec:eigenvariance_interference}), incoherence (region A, \S\ref{subsubsec:eigenvariance_incoherent}), feathering (region C, \S\ref{subsubsec:eigenvariances_feathering}), or single power spectrum (region D, \S\ref{subsubsec:eigenvariances_singlepk}). Only the first mechanism, interference, applies to $B_{12}$, while all four mechanisms occur in $B_{23}$ and $B_{31}$. 

The incoherence, feathering, and single power spectrum mechanisms arise from differences in the rate at which $k_3$ varies with $k_1$ across a configuration. At fixed $\delta$ and $\theta$, the wavenumbers $k_2$ and $k_3$ vary with $k_1$ according to equation \pref{eqn:k2_k3_fcn_delta_theta}; their derivatives with respect to $k_1$ at fixed $\delta$ and $\theta$ are
\begin{align}
\label{eqn:dk2_dk1_dk3_dk1}
\frac{\mathrm{d}k_2}{\mathrm{d}k_1} &= 1,\nonumber\\
\frac{\mathrm{d}k_3}{\mathrm{d}k_1} &= \frac{(k_1 + k_2)(1 + \cos{\theta})}{k_3} \nonumber\\
& = \frac{(2k_1 + \delta \lambda_f / 2)(1 + \cos{\theta})}{\sqrt{k_1 \left(1+\cos{\theta}\right)\left( 2k_1 + \delta \lambda_f \right) + \left(\delta \lambda_f/2 \right)^2}}.
\end{align}
The behavior of $dk_3/dk_1$ differs across the three regions marked in the right two panels of Figure~\ref{fig:k31_term_regions_map}. First, at the left edge of the RMS map (region B) where $\theta=0$, $k_1$ and $k_2$ are parallel, so $k_3=k_1+k_2$ and $\cos{\theta}=1$. The derivative in equation \pref{eqn:dk2_dk1_dk3_dk1} then simplifies to $dk_3/dk_1 = 2$. Second, in regions C and D where $\theta$ approaches $\pi$, $\bk_2$ is antiparallel to $\bk_1$, so $k_3$ is the difference between the other two sides: $k_3 = k_2-k_1 = \delta$. In other words, as $\theta \to \pi$ in these configurations, $dk_3/dk_1 \to 0$, and $k_3$ is independent of $k_1$ for any fixed $\delta$ and $\theta$ configuration. Third, the only configuration where $dk_3/dk_1$ is unity for all $k_1$ is the equilateral triangle in region A, where $\theta/\pi=2/3$ and $\delta=0$, implying $k_1=k_2=k_3$. In general, the rate of change of $k_3$ with $k_1$ increases as $\theta$ decreases or $\delta$ increases. 

In regions approaching $\theta=0$ or $\theta/\pi=1$, therefore, $k_3$ may vary across a configuration twice as quickly as $k_1$, or not at all. When shown as a function of $k_1$, the oscillations in $\PBAO_3$ are consequently stretched (as $\theta \to \pi$) or compressed (as $\theta \to 0$) relative to the oscillations in $\PBAO_1$. While the wavelength of oscillations in $\PBAO_1$ and $\PBAO_2$ is the BAO fundamental wavelength $\lambda_f$ (equation~\ref{eqn:def_lambda_f}), the wavelength of oscillations in $\PBAO_3$ can be infinitely large or as small as $\lambda_f/2$. We find that the interference picture is a good description of the interaction between two oscillations when the ratio of their wavelengths is less than roughly 1.4; when the wavelength of $\PBAO_3$ differs from that of $\PBAO_2$ and $\PBAO_1$ by more than this factor, the concept of a phase shift between $\PBAO_3$ and the other power spectrum in the product becomes meaningless because the wavelengths are simply too different. In the products $\PBAO_2\PBAO_3$ and $\PBAO_3\PBAO_1$, then, the BAO amplitude is no longer determined by any phase shift between the oscillations, but instead by the alignment of the first (and highest, as subsequent peaks will be suppressed by Silk damping) peaks in each. 

\subsubsection{Interference}
\begin{figure*}
\includegraphics{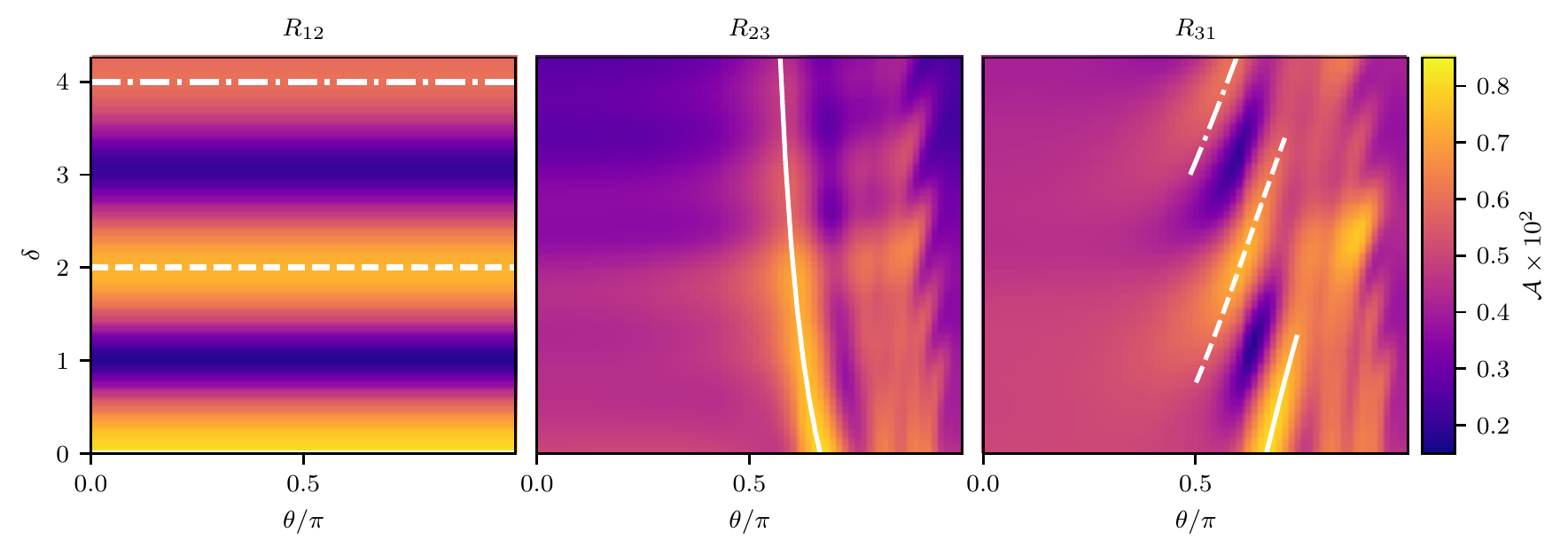}
\caption{\label{fig:eigenplots_single} When two power spectra are in phase---that is, when $k_i$ and $k_j$ differ by a multiple $n$ of the BAO fundamental wavelength $\lambda_f$---constructive interference increases $\mathcal{A}$, as discussed further in \S\ref{subsec:single_term_variance}. The curves show $k_2=k_1+n\lambda_f$ (left panel, where $n=\delta/2$ as odd integer values of $\delta$ produce destructive interference), $k_3=k_2+n\lambda_f$ (middle panel), and $k_3=k_1+n\lambda_f$ (right panel). Solid curves have $n=0$, dashed $n=1$, and dot-dashed $n=2$. For $R_{23}$ and $R_{31}$ (middle and right), the curves are calculated assuming $k_1=0.1 \Mpch$, i.e., in the middle of the $k_1$ range used in this work. Curves are shown only where the wavelength of $\PBAO_i$ as a function of $k_1$ differs from the wavelength of $\PBAO_1$ by less than a factor of 1.4 (as explained in \S\ref{subsec:single_term_variance}), which in $R_{23}$ is the case only for $n=0$. At higher and lower $\theta$ where the wavelengths differ more widely, the RMS amplitude is driven not by the phase difference between the two spectra, but instead by the alignment of individual peaks, as described in \S\ref{subsubsec:eigenvariance_incoherent}, \S\ref{subsubsec:eigenvariances_feathering}, and \S\ref{subsubsec:eigenvariances_singlepk}.}
\end{figure*}

Our basis was designed to highlight interference effects between pairs of power spectra, which determine $\mathcal{A}$ in the regions marked ``A'' of Figure~\ref{fig:k31_term_regions_map}. As outlined in \S\ref{sec:interferometric_basis}, when two wavenumbers $k_i$ and $k_j$ differ by a multiple of the fundamental BAO wavelength $\lambda_f$ (equation \ref{eqn:def_lambda_f}), the two power spectra $\PBAO_i$ and $\PBAO_j$ interfere constructively and amplify BAO. In all three terms shown in Figure~\ref{fig:eigenplots_single}, interference produces bright ridges of amplified BAO corresponding to the configurations given by equations  (\ref{eqn:k1_k2_equal_delta}--\ref{eqn:k1_k3_equal_delta}), where two wavenumbers differ by $n\lambda_f$.

\label{subsubsec:eigenvariance_interference}

The left panel of Figure~\ref{fig:eigenplots_single} is clearly similar to the low-$\theta$ region of the full RMS map (Figure~\ref{fig:full_colorplot}) where $B_{12}$ dominates. RMS amplitude is maximized for $\delta=0$, where $k_1=k_2$ and the two BAO features are perfectly in phase. Constructive interference repeats wherever the phase difference between the two power spectra is a multiple of the BAO fundamental wavelength $\lambda_f$, that is, for even integer values of $\delta$. The first two harmonics are marked in the left panel of Figure~\ref{fig:eigenplots_single}. As $\delta$ increases, the maximum $\mathcal{A}$ at even $\delta$ declines. This is a result of the declining amplitude of the BAO feature at small scales due to Silk damping. As $\delta$ rises, the $k_1$ dependence is unchanged, but $k_2$ becomes large enough that Silk damping reduces the amplitude of $\PBAO_2$; as the BAO wiggle contribution is damped, it provides less enhancement. In practice, nonlinear structure formation would also degrade the BAO signal at large $\delta$, similar to the effects at large $k_1$ discussed in \S\ref{sec:interferometric_basis}.

Power spectra also interfere constructively and destructively to produce distinct ridges and troughs in the $R_{23}$ and $R_{31}$ RMS maps (right two panels of Figure~\ref{fig:eigenplots_single}). As in $R_{12}$, we expect the RMS amplitude in $R_{23}$ to be maximized when $\PBAO_2$ and $\PBAO_3$ are in phase or differ by a multiple of the wavelength, and the RMS amplitude in $R_{31}$ to be maximized when $\PBAO_3$ and $\PBAO_1$ are in phase or differ by a multiple of $\lambda_f$. The solutions (equations \ref{eqn:k2_k3_equal_delta} and \ref{eqn:k1_k3_equal_delta}) to $k_2=k_3 + n \lambda_f$ and $k_3=k_1 + n \lambda_f$, however, depend not only on $\delta$ and $\theta$, but also on $k_1$. As a result, for a single choice of $(\delta,\theta)$, it is not possible for $k_2$ to equal $k_3$ (or $k_3$ to equal $k_1$) for all $k_1$ in a configuration. We therefore choose $k_1=0.1 \Mpch$ (that is, in the middle of our $k_1$ range) as a representative value of $k_1$. We evaluate equation \pref{eqn:k2_k3_equal_delta} at $k_1=0.1 \Mpch$ to compute the curve of $k_3=k_2$ shown in the middle panel of Figure~\ref{fig:eigenplots_single}.
This curve does correspond to maximum $\mathcal{A}$, but the $k_3=k_2+\lambda_f$ curve does not; it falls in region B (labeled in Figure~\ref{fig:k31_term_regions_map} and discussed in \S\ref{subsubsec:eigenvariance_incoherent}), where the wavelengths of $\PBAO_3$ and $\PBAO_2$ are widely different and the interference picture no longer applies. 
We also evaluate equation \pref{eqn:k1_k3_equal_delta} at $k_1=0.1 \Mpch$ to compute the curves of $k_1=k_3 + n \lambda_f$ shown in the right panel of Figure~\ref{fig:eigenplots_single}.

\subsubsection{Incoherence}
\label{subsubsec:eigenvariance_incoherent}
In Region B (labeled in Figure~\ref{fig:k31_term_regions_map}) of the $R_{23}$ and $R_{31}$ RMS maps, the RMS amplitude is relatively uniform; $\mathcal{A}$ is neither maximized nor minimized for these configurations. Because the wavelength of $\PBAO_3$ is much shorter than that of the other two power spectra in this region, the power spectra entering the products $\PBAO_2\PBAO_3$ and $\PBAO_3\PBAO_1$ are incoherent: they cannot interfere constructively or destructively, and patterns in $\mathcal{A}$ arise from the amplitudes of the largest-scale peaks in the power spectra. BAO amplitude can only be enhanced when two peaks---a single pair---in the two power spectra align with and amplify each other, and the greatest amplitude occurs where these peaks are at large scales and therefore minimally Silk-damped.

\begin{figure}
\includegraphics{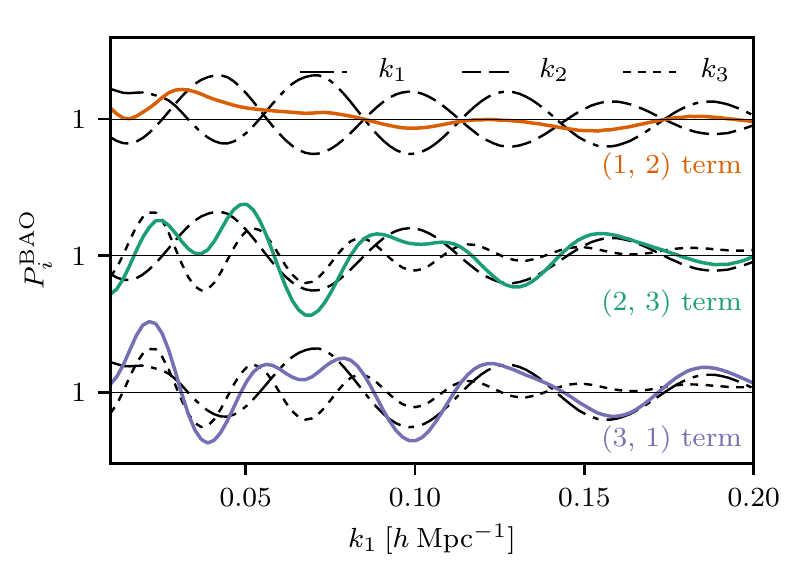}
\caption{\label{fig:interference_low_theta} For a configuration with $\theta/\pi=0.2$ and $\delta=1$, the RMS amplitude in the $R_{12}$ (uppermost set of curves above) term is driven by phase differences (i.e., interference, described in \S\ref{subsubsec:eigenvariance_interference}), while the pattern in the $R_{23}$ (middle set) and $R_{31}$ (lower set) terms is a result of wavelength differences (i.e., incoherence, described in \S\ref{subsubsec:eigenvariance_incoherent}). Black curves show the ratio of the linear to the no-wiggle power spectrum, $\PBAO_i = P(k_i)/P_{\rm nw}(k_i)$, for each wavenumber as it varies with $k_1$; the product of each pair of ratios is shown in color ($\PBAO_1\PBAO_2$ in orange, $\PBAO_2\PBAO_3$ in teal, and $\PBAO_3\PBAO_1$ in lavender). For example, the oscillations in $\PBAO_1$ and $\PBAO_2$ are out of phase, so the power spectra interfere destructively in $P_{12}$ (orange, discussed in \S\ref{subsubsec:eigenvariance_interference}). In contrast, $P_{23}$ and $P_{31}$ include $\PBAO_3$. At low $\theta$, $k_3$ can vary over more than twice the range of $k_1$, so the oscillations in $\PBAO_3$ are compressed relative to the others (e.g., compare the short-dashed curve to the dot-dashed curve in the lower set of curves). As can be seen in the lavender $P_{31}$ term, the two interfering oscillations have very different wavelengths, so their product is neither ``constructive'' nor ``destructive.'' The behavior in $P_{23}$ (teal) is similar; see \S\ref{subsubsec:eigenvariance_incoherent} for further discussion. Figure is reproduced from \protect\cite{2018arXiv180611147C}.} 
\end{figure}

As $\theta \to 0$, each configuration spans a wider range of $k_3$ for a fixed range of $k_1$, meaning that any change in $k_1$ maps to a larger change in $k_3$ (equation \ref{eqn:dk2_dk1_dk3_dk1}). In the $\theta=0$ limit, for example, $k_3=k_1+k_2$ spans at least twice the range of $k_1$. The wavelength of $\PBAO_3$ as a function of $k_1$ is compressed relative to that of $\PBAO_2$ as a function of $k_1$ ($k_2$ everywhere changes with $k_1$ at the same rate, since these two are related by addition of $\delta$). In the small-$\theta$ region, therefore, all the products of power spectra in terms that include $k_3$---$R_{23}$ and $R_{31}$---are products of oscillations with very different wavelengths (see Figure~\ref{fig:interference_low_theta}). 
\label{subsec:13_or_23_regionA}

Although most of the pattern is washed out in the low-$\theta$ region of the $R_{23}$ RMS map (middle panel of Figure~\ref{fig:eigenplots_single}), faint banding is still visible around integer values of $\delta$. The maxima diminish with increasing $\delta$ as they do in $R_{12}$---the amplitude of the BAO oscillation in the power spectrum drops at higher wavenumbers.  The banding is a result of the relative phase (controlled by $\delta$) between $\PBAO_2$ and $\PBAO_3$ at low $k_1$. Because the wavelengths of $\PBAO_2$ and $\PBAO_3$ are very different, the peaks do not align more than once. The RMS amplitude is highest when the pair of aligned peaks are both large, but the amplitude of the oscillation in each $\PBAO$ falls with increasing wavenumber. Therefore, BAO are maximized when the lowest-wavenumber peak (or trough) in $\PBAO_2$ aligns with the lowest-wavenumber peak (or trough) in $\PBAO_3$. When the lowest-wavenumber peak in $\PBAO_2$ aligns with a trough in $\PBAO_3$, the small-wavenumber contributions cancel, and BAO are minimized. 

The $\theta \to 0$ region of the $R_{31}$ RMS map behaves similarly to the same region in the $R_{23}$ RMS map---the wavelength of $\PBAO_3$ is again much shorter than the wavelength of $\PBAO_1$. The phase of $\PBAO_1$ is fixed, so the faint banding pattern is diminished in $R_{31}$. Silk damping still decreases the amplitude of oscillations in $\PBAO_3$ as $\delta$ increases; at large $\delta$, $\PBAO_3$ becomes smooth and approaches unity. The RMS amplitude in $R_{31}$ therefore approaches that of $\PBAO_1$ only. Near $\delta=2$, the small-wavenumber region is a minimum of $\PBAO_3$. As in $R_{12}$, contribution of the low-wavenumber region to the amplitude is therefore minimized, resulting in a faint minimum in the RMS map. 

\subsubsection{Feathering}
\label{subsubsec:eigenvariances_feathering}
\begin{figure}
\includegraphics{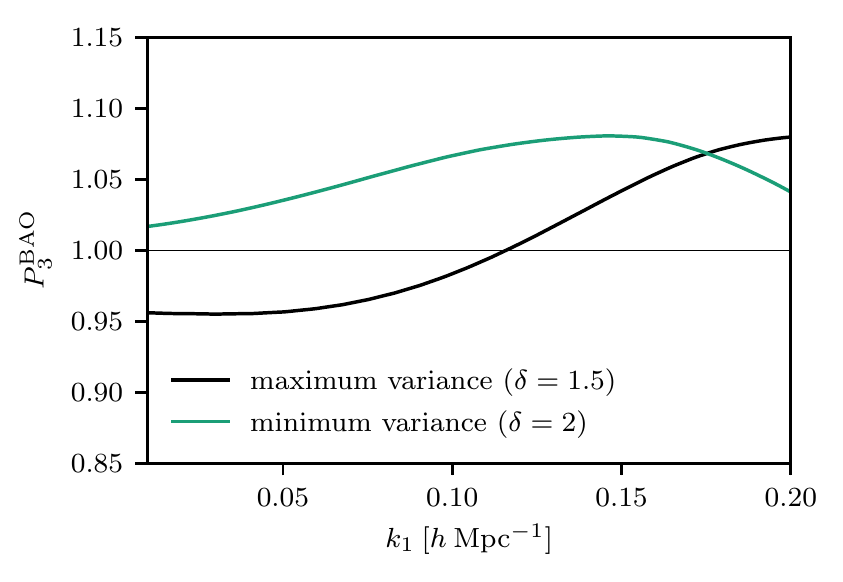}
\caption{\label{fig:feathering_diagram} For configurations in the ``feathering'' region (Region C of Figure~\ref{fig:k31_term_regions_map}), BAO amplitude is driven by the long-wavelength oscillation in $\PBAO_3$. The BAO amplitude $\mathcal{A}$ is maximized when $\PBAO_3$ varies fully, from trough to peak (black), and minimized when $\PBAO_3$ covers only half of that range (green). See \S\ref{subsubsec:eigenvariances_feathering} for further discussion.}
\end{figure}

In Region C of the $R_{23}$ and $R_{31}$ RMS maps in Figure~\ref{fig:k31_term_regions_map}, small maxima and minima alternate as $\delta$ increases. We refer to this behavior as ``feathering,'' a pattern of bright feathers alternating with regions of lower $\mathcal{A}$. As in Region B (\S\ref{subsubsec:eigenvariance_incoherent}), the behavior of $\mathcal{A}$ in this region results from the difference in wavelength between $\PBAO_3$ and the other two power spectra. In Region C, $\theta$ approaches $\pi$, so $dk_3/dk_1$ (equation \ref{eqn:dk2_dk1_dk3_dk1}) is small and $k_3$ changes little with $k_1$. $P_{31}$ is then $\PBAO_1$ modulated by a stretched-out and slowly varying $\PBAO_3$. Across the full range of $k_1$, $\PBAO_3$ traverses half a wavelength. If this half wavelength starts from an extremum of $\PBAO_3$ where $k_1$ is small, and ends at the other extremum of $\PBAO_3$ where $k_1$ is large (see Figure~\ref{fig:feathering_diagram}), $\mathcal{A}$ is maximized. In contrast, $\mathcal{A}$ is minimized between the bright feathers, where instead $\PBAO_3 \sim 1$ for small $k_1$, and again $\PBAO_3 \sim 1$ for the highest $k_1$ in our range. In this case, the range of $\PBAO_3$ is halved relative to the maximum case, and the amplitude contribution due to the $k_3$ modulation is minimized. 
\label{subsec:13_or_23_regionC}

In region C of the $R_{31}$ RMS map (right panel of Figure~\ref{fig:k31_term_regions_map}), bright feathers alternate with brighter feathers (for example, the maximum at $\delta=2.5$, $\theta/\pi=0.89$). This alternating pattern arises as $\PBAO_3$ at small $k_1$ moves from a trough, to unity, to a peak. Because Silk damping reduces BAO amplitude at small scales, $\PBAO_1$ is maximized for small $k_1$. If this maximum coincides with a maximum in $\PBAO_3$, its contribution to the RMS amplitude is larger than when it coincides with a minimum in $\PBAO_3$. That is, $\mathcal{A}$ is greater when $\PBAO_3$ travels from a peak to a trough than vice versa, because Silk damping reduces the contribution of peaks at high $k_1$ that coincide with the final peak in the latter case. Therefore, while all bright feathers occur where $\PBAO_3$ starts from an extremum, they are brighter where that extremum is a maximum and dimmer where it is a minimum. As $\delta$ increases, $\mathcal{A}$ declines for the feathers, for the same reason as the $R_{12}$ interference described in \S\ref{subsubsec:eigenvariance_interference}. At high $\delta$, $k_3$ is larger, so Silk damping reduces the amplitude of oscillations. 

Similar logic holds for $R_{23}$ (middle panel of Figure~\ref{fig:k31_term_regions_map}). Bright feathers occur where $\PBAO_3$ is either a peak or a trough at low $k_1$, and the opposite at high $k_1$; $\mathcal{A}$ is minimized where instead $\PBAO_3$ is unity at both small and large wavenumber. However, unlike $R_{31}$, there is no pattern of alternating brighter and dimmer maxima. In $R_{31}$, $\PBAO_1$ is held fixed while the starting point of the oscillation in $\PBAO_3$ varies, so the initial peak in $\PBAO_1$ can correspond to either a trough in $\PBAO_3$ (as in Figure~\ref{fig:feathering_diagram}) or a peak. But in $R_{23}$, the starting points of both $\PBAO_2$ and $\PBAO_3$ depend on $\delta$---and differ by $k_1$. In this region the magnitude of $k_3$ is determined by the difference $k_2-k_1$. At the initial value of $k_1$ in our range, $k_1 = 0.01 \hMpc$, the magnitude of $k_3$ is roughly $k_2 - 0.01 \hMpc$. $k_2$ and $k_3$ therefore differ by less than one-fourth of the BAO fundamental wavelength $\lambda_f$. Whether $\PBAO_3$ is maximized or minimized at $k_1=0.01 \hMpc$, $\PBAO_2$ at $k_1=0.01 \hMpc$ must fall in the same quarter wavelength, so its value must be close to that of $\PBAO_3$ but closer to unity. Since the difference between $k_2$ and $k_3$ is fixed at small $k_1$ (unlike the difference between $k_1$ and $k_3$ at small $k_1$), an initial peak or trough in $\PBAO_3$ can only correspond to a more limited range of values of $\PBAO_2$, removing the mechanism by which the brightest feathers appear in the RMS map for $R_{31}$.

\subsubsection{Single Power Spectrum}
\label{subsubsec:eigenvariances_singlepk}
Configurations in Region D (labeled in Figure~\ref{fig:k31_term_regions_map}) are squeezed, and only one power spectrum term contributes BAO. With only one oscillation, there can be no interference to amplify BAO, so $\mathcal{A}$ is fairly uniform in Region D of the $R_{23}$ and $R_{31}$ RMS maps.

When $\theta/\pi = 1$, equation \pref{eqn:dk2_dk1_dk3_dk1} gives $dk_3/dk_1=0$; that is, $k_3$ is independent of $k_1$ for any choice of $\delta$. $R_{23}$ is then simply the oscillation from $\PBAO_2$ alone, with no interference. Similarly, $R_{31}$ reduces to $\PBAO_1$. In both terms, the bispectrum BAO come solely from the oscillation of one $\PBAO$. This oscillation is multiplied by $\PBAO_3$, which does introduce a slight dependence on the parameter $\delta$. While  $\PBAO_3$ is constant as a function of $k_1$ for any value of $\delta$, the value of that constant does depend on $\delta$: as in equation \pref{eqn:k3_thetapi_limit}, $k_3 = \delta \lambda_f/2$ for all $k_1$. The argument of $\PBAO_3$ changes with $\delta$, so the level of $\PBAO_3$ oscillates up and down as $\delta$ increases. In $R_{31}$, $\mathcal{A}$ along the $\theta/\pi=1$ line depends only on the level of $\PBAO_3$. When $\PBAO_3 > 1$, the entire oscillation in $\PBAO_1$ is stretched vertically by a factor greater than unity, slightly increasing $\mathcal{A}$. The opposite is true when $\PBAO_3 < 1$, which compresses the amplitude of the $\PBAO_1$ oscillation and decreases the RMS amplitude. This effect diminishes at higher $\delta$, as $\PBAO_3$ converges to unity. 

In $R_{23}$, again the oscillation of $\PBAO_3$ with changing $\delta$ causes the $\theta/\pi=1$ RMS amplitude to depend on $\delta$. Additionally, as $\delta$ increases, $\PBAO_2$ is increasingly Silk-damped, smoothly decreasing the RMS amplitude in $\PBAO_2$. The faint banding in both $R_{31}$ and $R_{23}$ is visible along the rightmost edge of the middle and right panels of Figure~\ref{fig:k31_term_regions_map}.

\subsection{Double Dominance}
\label{subsec:double_dominance_variance}
In the purple ($k_2>k_1$ but $k_3 \approx k_2$) and green ($\theta/\pi \sim 1$ and $\delta$ small) regions of the dominance map (Figure~\ref{fig:dominance_plot_guide}), two terms are of comparable magnitude. In these regions we calculate the RMS amplitude \pref{eqn:mathcalA}, shown in the second panel of Figure~\ref{fig:eigenplots_one_negligible}, of the ratio of the sum of the two dominant terms to its no-wiggle analog \pref{eqn:R_ijjkdef}.

\subsubsection{$B_{12}$ and $B_{31}$ Dominant}
\begin{figure}
\includegraphics{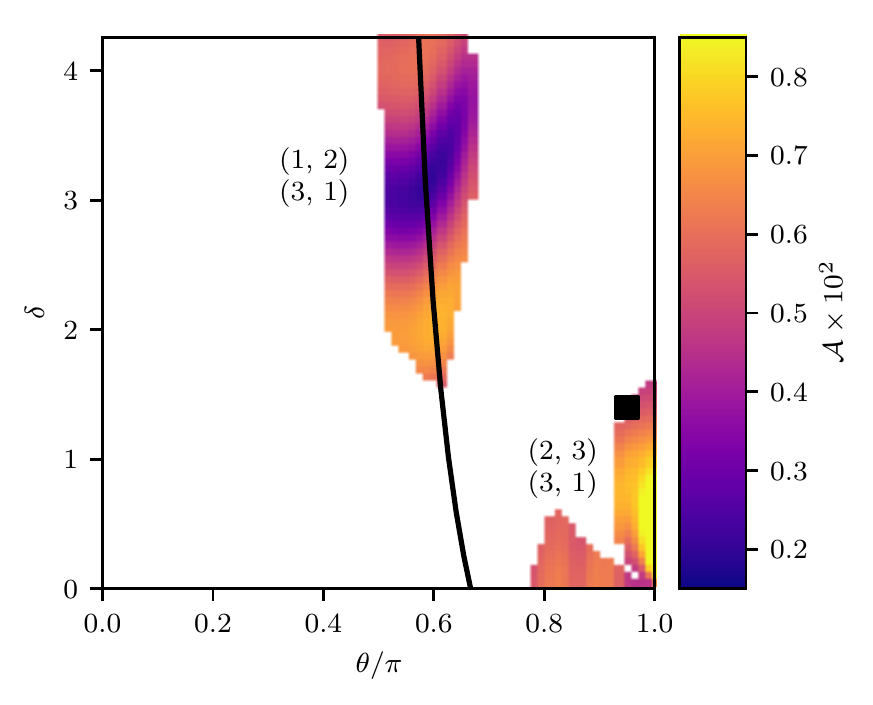}
\caption{\label{fig:doubleterms_k2_k3_equalline} Regions where two terms dominate the bispectrum (same as second panel of Figure~\ref{fig:eigenplots_one_negligible}) are discussed in \S\ref{subsec:double_dominance_variance}. The curve $k_2=k_3$ for $k_1=0.1\hMpc$ is shown in black. At $\theta$ to the left of this curve, the RMS map behaves like that for $R_{12}$, while for higher $\theta$, it is more similar to that for $R_{31}$. In both $R_{12}$ and $R_{31}$ in this region, $\mathcal{A}$ is determined by the interference mechanism of \S\ref{subsubsec:eigenvariance_interference}. As $\delta$ increases above the black square symbol, only $B_{31}$ dominates the cyclic sum.}
\end{figure}

The purple region of the dominance map (Figure~\ref{fig:dominance_plot_guide}), where $B_{12}$ and $B_{31}$ are both dominant, is a region of transition between $B_{12}$ dominance at smaller $\theta$ and $B_{31}$ dominance at higher $\theta$. The curve of $k_2=k_3$ passes through the center, as shown in Figure~\ref{fig:doubleterms_k2_k3_equalline}. Along this curve, $B_{12}$ and $B_{31}$ are very similar (but not identical, since the fact that $k_2$ is equal to $k_3$ for our representative $k_1=0.1\hMpc$ does not imply that $k_2=k_3$ for all $k_1$ in a configuration). At $\theta$ lower than the cutoff defined by the curve of $k_2=k_3$ in Figure~\ref{fig:doubleterms_k2_k3_equalline}, $B_{12}$ begins to dominate. While $B_{31}$ is still large, $k_3$ is close to $k_2$, so the oscillations and interference behavior of the two terms are very similar. The RMS amplitude $\mathcal{A}$ is maximum on the lines where $k_2 = k_1 + n \lambda_f$. The reverse holds at $\theta$ higher than the $k_2=k_3$ curve, where $B_{31}$ grows to become dominant. Again, the oscillatory behavior of the two terms is similar, with $B_{31}$ becoming dominant as $\theta$ continues to grow.

\subsubsection{$B_{23}$ and $B_{31}$ Dominant}
In the green region of the dominance map (Figure~\ref{fig:dominance_plot_guide}), $B_{23}$ and $B_{31}$ are dominant and $\mathcal{A}$ is maximized, as shown in the RMS map (Figure~\ref{fig:full_colorplot}). These are the squeezed configurations: $k_2$ points back along $k_1$, and is slightly longer by $\delta \lambda_f/2$, so the magnitude of $k_3$ is constant for any configuration---that is, when $\theta/\pi=1$ and $\delta$ is fixed, $k_3 = \delta \lambda_f/2$ for all $k_1$ as in equation \pref{eqn:k3_thetapi_limit}. The RMS amplitude arises from the sum of $B_{31}$, which is large and positive, and $B_{23}$, which is large and negative (as discussed in \S\ref{subsubsec:dominance_green}). Because $k_3$ is constant for all $k_1$ in a configuration, $\PBAO_3$ is a constant, so $R_{31} = \PBAO_3\PBAO_1 \propto \PBAO_1$ and $R_{23} = \PBAO_2\PBAO_3 \propto \PBAO_2$. In the ratio of the sum to its no-wiggle analog, $R_{23+31}$ (equation~\ref{eqn:R_ijjkdef}), oscillations arise from the difference between the $B_{23}$ and $B_{31}$ contributions: a sine added to a negative sine, only slightly out of phase. As $\delta$ is positive, the negative $B_{23}$ is always slightly smaller in magnitude than $B_{31}$, so $R_{23+31}$ remains positive.

As $\delta$ increases, $B_{23}$ shrinks and $B_{31}$ becomes dominant near the square symbol in Figures~\ref{fig:dominance_plot_guide} and \ref{fig:doubleterms_k2_k3_equalline}, as discussed in \S\ref{subsubsec:dominance_green}. 
The RMS map transitions into the $B_{31}$-dominant region described in \S\ref{subsubsec:eigenvariances_singlepk} above where only one power spectrum contributes BAO to the bispectrum.

\subsection{No Term Negligible}
\label{subsec:eigenvariance_no_term_negligible}
In the final region around equilateral triangles (black region, Figure~\ref{fig:dominance_plot_guide}), all three sides are comparable and no term can be disregarded. We must simply reproduce the calculation of the full RMS map in this region, as shown in the middle right panel of Figure~\ref{fig:eigenplots_one_negligible}. The RMS amplitude $\mathcal{A}$ is maximized at the equilateral triangle ($\delta=0, \, \theta/\pi=2/3$), where all three sides of a triangle are equal---and therefore the $\Ftwo$ kernels are all equal and the power spectra are all in phase.

\section{Discussion}
\label{sec:discussion}
\subsection{Implications for the Reduced Bispectrum}
\label{subsec:reduced}
For many triangle configurations, we find that the full bispectrum RMS map is described well by the behavior of only one or two terms in the cyclic sum. The large dynamic range of the $\Ftwo$ kernel can separate terms by an order of magnitude, allowing us to disregard smaller terms when computing the RMS amplitude. If this held for the reduced bispectrum (defined in e.g. \citealt{2002PhR...367....1B})
\begin{equation}
\label{eqn:reduced_bispec_Q}
Q(k_1, k_2, k_3) = \frac{B(k_1, k_2, k_3)}{P(k_1)P(k_2) + P(k_2)P(k_3) + P(k_3)P(k_1)},
\end{equation}
we could simplify its behavior as well. Our work does show that there are regions where the numerator of $Q$ can indeed be simplified. 

However, a more useful approximation would be if one could approximate the denominator by just one term rather than the full cyclic sum of power spectra, or even a pair of products. Unfortunately, though, it is primarily the $\Ftwo$ kernel that drives the dominance of one term relative to the others (see \S\ref{subsec:f2_kernel_vs_pk_dominance}), and it does not enter the denominator of $Q$. As the leftmost panel of Figure~\ref{fig:dominant_terms_p_f2} shows, a large swatch of the $\delta$-$\theta$ plane is black (no term negligble) for the relevant products of power spectra $P_{ij}$. While there are several regions where two terms dominate the others (purple, blue, green), these do not seem to offer a significant simplification as they still produce a complicated denominator in equation~\pref{eqn:reduced_bispec_Q}. 

It is only in the red ($B_{12}$ dominant) and blue ($B_{31}$ dominant) regions that the denominator greatly simplifies, to respectively $P_{12}$ or $P_{31}$. Since the dominant term in $\Ftwo$ is always the same as that in $P_{ij}$ (see \S\ref{subsec:ordering_of_subdominant_terms} and Figure~\ref{fig:max_min_term_differs}), in these regions $\Ftwo_{12}$ and $\Ftwo_{31}$, respectively, will be much larger than the other two $\Ftwo_{ij}$. The bispectrum is therefore well-approximated by respectively $\Ftwo_{12} P_{12}$ and $\Ftwo_{31} P_{31}$. Thus, in these limited regions, $Q$ reduces to $\Ftwo_{12}$ and $\Ftwo_{31}$. In short, working in the $\delta$-$\theta$ basis does highlight convenient triangles where one can directly measure the growth kernel $\Ftwo$ alone and easily divide out the linear theory density field statistics. The contribution of gravitational growth can thereby be isolated from that of the linear theory density field. This isolation might be especially useful in using the 3PCF as a probe of modified gravity (e.g., Vernizzi et al. 2018, in prep.).

Especially insofar as high-wavenumber details of the power spectrum sourced by baryon physics remain challenging to model, canceling out the power spectrum from measurements of $\Ftwo$ may be desirable. Of course this must be weighed against the reduction in number of usable configurations, as this cancellation happens only on limited regions of the $\delta$-$\theta$ plane. Further, at the wavenumbers where baryons become relevant, a tree-level, linearly biased model of the bispectrum is likely already beginning to falter; the numerator is measuring higher-order perturbation theory kernels and higher-order biasing even in these ``simpler,'' single-dominance regions.

\subsection{Connection to Real Space}
We now briefly discuss the connection of the present paper to the 3PCF in configuration space (i.e., real space without redshift-space distortions). \cite{Hoffmann:2017kml} further discuss differences and similiarites between bispectrum and 3PCF more generally, though with a focus on bias parameters, most relevant for smaller scales than the BAO scales investigated here. The wiggles in the bispectrum ultimately correspond to sharp features in configuration space, in particular the BAO creases where one triangle side is the BAO scale or twice the BAO scale. These are visible in Figure 7 of \cite{Slepian:2016weg}, particularly in the linear bias ($\ell=1$) panel but also more faintly in $\ell=0$ and $\ell=2$. The intuition is much the same as with the 2PCF and power spectrum, where a bump in configuration space leads to a harmonic series of oscillations in Fourier space. One important difference here is that the $\Ftwo$ kernel, which weights the products of power spectrum by $k^{\pm 1}$, acts like a derivative in configuration space. The BAO feature in the 3PCF is thus essentially the derivative of a BAO bump: positive as the BAO bump rises, then zero at the BAO scale, and negative at larger separations as the BAO bump falls.

\subsection{Simplification of Multipole Basis}
We now point out an interesting additional implication of our work. The multipole basis (expanding the angular dependence of the 3PCF or bispectrum in Legendre polynomials), proposed in \cite{2004ApJ...605L..89S} (see also \citealt{Pan:2005ym}), has recently been exploited in a series of works \citep{Slepian:2015qza, Slepian:2015qwa, 2017arXiv170900086F} to accelerate measurement of the 3PCF. However, in practice that approach truncates the multipole expansion of the 3PCF as one measures it. The works cited above chose a maximum multipole of $\ell_{\rm max} = 10$. In principle, however, even at tree level the 3PCF has support out to infinite $\ell$, as the expansion is done with respect to $\hat{\bk}_1\cdot\hat{\bk}_2$ and $k_3$ and $1/k_3$ have an infinite multipole series in this variable. 

In practice the 3PCF seems well-converged when summed into a function of opening angle using different numbers of multipoles (see Figure~8 of \citealt{Slepian:2015qza}). Our work shows that for certain configurations the multipole support is in principle finite. In the regions dominated by $B_{12}$, the bispectrum multipole expansion has compact support, requiring (at least at tree level) only $\ell=0, 1,$ and $2$. The same will hold for the 3PCF; \cite{2004ApJ...605L..89S} shows that a given $\ell$ in Fourier space maps to only the same $\ell$ in configuration space. (This immediately follows from the plane wave expansion into spherical harmonics and spherical Bessel functions, use of the spherical harmonic addition theorem, and orthogonality of the spherical harmonics.) 

There are two implications here: first, the adequacy of tree level perturbation theory can be easily tested using a very small set of multipoles in the red region of $B_{12}$ dominance. Second, within this restricted region, the computational work and covariance matrix dimension can be greatly reduced by measuring the 3PCF in the multipole basis only to $\ell_{\rm max} = 2$. Of course, the price is the reduced number of configurations (and signal) available. While this level of compression may not be necessary for isotropic statistics, RSD introduce a much richer angular structure at a fixed $\ell_{\rm max}$ (see \citealt{Slepian:2017lpm, 2018arXiv180302132S}), so a reduction in $\ell_{\rm max}$ may be of particular value.

\section{Conclusion}
\label{sec:conclusion}
Our bispectrum basis (\S\ref{sec:interferometric_basis}), designed to identify triangle configurations that amplify the BAO signal, also provides insight on the structure of BAO in the bispectrum. Our analysis in \S\ref{sec:dominance_regions} shows that for certain triangle shapes, the bispectrum is dominated by only one or two terms of the cyclic sum~\pref{eqn:B0}. The dominance structure is driven primarily by the $\Ftwo$ kernel of Eulerian standard perturbation theory (\S\ref{subsec:f2_kernel_vs_pk_dominance}), which is highly sensitive to triangle shape. In \S\ref{sec:analytic_eigen} we show analytically that, because BAO are a small feature relative to the broadband bispectrum, the RMS BAO amplitude in the full bispectrum reduces to the RMS BAO amplitude in the dominant term or terms. The error in this approximation is suppressed by one order relative to the BAO amplitude itself. In \S\ref{sec:eigenvariances}, we build up the complete RMS map of the dependence of BAO amplitude on triangle parameters from simpler maps. These maps show the RMS BAO amplitude in each of the three terms contributing to the cyclic sum. In regions where the corresponding terms dominate the cyclic sum, the full RMS BAO amplitude is well approximated by the single-term-dominant or double-term-dominant maps. We reproduce the full bispectrum RMS map by stitching together these simpler maps in the regions where they provide the dominant contribution to bispectrum BAO, then fully discuss the mechanisms that drive BAO amplitude in each single-term-dominant (\S\ref{subsec:single_term_variance}) and double-term-dominant (\S\ref{subsec:double_dominance_variance}) RMS map. 

The BAO amplitude in each single term is determined by one of four mechanisms: interference (\S\ref{subsubsec:eigenvariance_interference}), incoherence (\S\ref{subsubsec:eigenvariance_incoherent}), feathering (\S\ref{subsubsec:eigenvariances_feathering}), or single power spectrum (\S\ref{subsubsec:eigenvariances_singlepk}). The first mechanism, interference, results from phase differences between two power spectra, and dramatically amplifies the BAO signal. The other three mechanisms occur where the wavelengths of the two BAO features are widely different, so the interaction between the two power spectra cannot consistently amplify BAO. 

Finally, in \S\ref{sec:discussion} we outline implications of our work for the reduced bispectrum, its connection to the 3PCF, and the potential to simplify the multipole expansion of the bispectrum and 3PCF for certain triangle shapes.

In a previous paper \citep{2018arXiv180611147C}, we used the interferometric basis detailed here to obtain substantial improvement in BAO constraints over the power spectrum alone, using a relatively small number of bispectrum measurements that carry the most BAO information. Ideally, bispectrum measurements on all possible triangles would be used to constrain the BAO scale. However, the number of mock catalogs needed to accurately estimate and invert the covariance matrix scales with the number of triangles \citep{Percival:2013sga}. The number of triangles that can be measured is therefore limited by the number of mock catalogs available, and bispectrum BAO constraints like those of \cite{2018MNRAS.478.4500P} are limited by the error in the covariance matrix. Since current resources limit the number of triangles that can be used to constrain BAO in the bispectrum, the best constraints will be obtained from the triangles that carry the most BAO information and are most independent from each other. One way to identify these triangles is by measuring the full covariance matrix, but such an approach faces the same initial problem of limited mock catalogs. 

We therefore face a circular problem: because it is computationally prohibitive to use fully $N$-body mocks to constrain the covariance matrix of all bispectrum triangles, we wish to reduce the size of the covariance matrix by selecting a subset of optimal triangles for BAO constraints. But without the full covariance matrix, how can those triangles be identified? Our basis offers a compression to only those triangles that are most sensitive to BAO, enabling a 15\% improvement over power spectrum BAO constraints using relatively few bispectrum measurements \citep{2018arXiv180611147C}. 

Of course, the optimal set of triangles for BAO measurement depends not only on the amplitude of the BAO signal in each configuration, but also on its signal to noise ratio and its covariance with previously measured configurations. In future work, we will further develop an algorithm for selecting triangle configurations, assuming the number of mock catalogs available limit the number of bispectrum measurements that can be used.

In future work, we will use also use BAO-sensitive triangles to better understand the covariance structure of BAO in the bispectrum and power spectrum. Reconstruction \citep{Eisensteinetal:07, 2009PhRvD..80l3501N, 2009PhRvD..79f3523P, Padmanabhanetal:12} is expected to affect the covariance between the power spectrum and bispectrum (\citealt{Schmittfull:2015mja}; see also \citealt{Slepian:2016nfb}, \S8.2), but as reconstruction is a numerical procedure, its effect on covariance is difficult to model analytically. Like the bispectrum measurements discussed in the previous paragraph, a full numerical study of the covariance between the post-reconstruction power spectrum and the pre-reconstruction bispectrum is limited by the number of fully $N$-body mocks available. Fewer mocks are needed if analysis is restricted to the set of triangles most sensitive to BAO, reducing the dimension of the covariance matrix. We will study the effects of reconstruction on these triangle configurations. This effort will allow us to combine bispectrum measurements with the post-reconstruction power spectrum. Depending on the level of independence, the combination of bispectrum measurements and reconstruction may offer further improvement in BAO constraints over that offered by reconstruction alone.

Our approach offers many further applications to the study of BAO in the bispectrum, which we plan to address in future work. For example, the phase of BAO in the power spectrum is sensitive to $N_{\rm eff}$, the effective number of relativistic neutrino species \citep{Bashinsky:2003tk, Follin:2015hya, Baumann:2017lmt, Baumann:2018qnt}. Our basis is very sensitive to phase effects, so it may be useful to constrain $N_{\rm eff}$ using the bispectrum (Child et al. 2019, in prep.). Other sources of a phase shift in power spectrum BAO such as relative velocities between baryons and dark matter \citep{2010JCAP...11..007D, 2010PhRvD..82h3520T, 2011JCAP...07..018Y, 2016PhRvL.116l1303B, 2016PhRvD..94f3508S}, constrained in the power spectrum by \cite{2013PhRvD..88j3520Y} and \cite{2017MNRAS.470.2723B} and in the 3PCF by \cite{Slepian:2016nfb}, may also be constrained using our interferometric basis. Last, our approach may enable study of massive spinning particles, which, if present during inflation, introduce oscillatory cosine terms in the bispectrum \citep{MoradinezhadDizgah:2018ssw}. These terms depend on the wavenumbers, so they can interfere with each other when cyclically symmed. 

\section*{Acknowledgements}
We thank Salman Habib, Uro\v{s} Seljak, and Martin White for useful discussions. ZS thanks Joshua Silver and Roberta Cohen for hospitality in Chicago. HC was supported by the National Science Foundation Graduate Research Fellowship Program under Grant No. DGE-1746045 and an international travel allowance through Graduate Research Opportunities Worldwide (GROW). Any opinions, findings, and conclusions or recommendations expressed in this material are those of the author(s) and do not necessarily reflect the views of the National Science Foundation. The work of HC  at  ANL  was  supported  under  the  U.S.  DOE contract DE-AC02-06CH11357. Support for the work of ZS was provided by the National Aeronautics and Space Administration through Einstein Postdoctoral Fellowship Award Number PF7-180167 issued by the Chandra X-ray Observatory Center, which is operated by the Smithsonian Astrophysical Observatory for and on behalf of the National Aeronautics Space Administration under contract NAS8-03060. ZS also received support from the Berkeley Center for Cosmological Physics (BCCP), and is grateful to both BCCP and Lawrence Berkeley National Laboratory for hospitality during this work.




\bibliographystyle{mnras}
\bibliography{refs}








\bsp	
\label{lastpage}
\end{document}